\documentclass[iop]{emulateapj}

\journalinfo{To Appear in ApJ}
\submitted{Received 2010 July 29; accepted 2011 April 3}

\shorttitle{MW Formation Through Halo Substructure. II.}
\shortauthors{Schlaufman et al.}

\begin{document}

\title{Insight Into the Formation of the Milky Way Through Cold Halo
Substructure. II. The Elemental Abundances of ECHOS}

\author{Kevin C. Schlaufman\altaffilmark{1,5},
Constance M. Rockosi\altaffilmark{1,6,7}, Young Sun Lee\altaffilmark{2},
Timothy C. Beers\altaffilmark{2}, and Carlos Allende Prieto\altaffilmark{3,4}}

\altaffiltext{1}{Astronomy and Astrophysics Department, University
of California, Santa Cruz, CA 95064, USA; kcs@ucolick.org and
crockosi@ucolick.org}
\altaffiltext{2}{Dept. of Physics and Astronomy and JINA: Joint Institute
for Nuclear Astrophysics, Michigan State University, E. Lansing, MI 48824,
USA; lee@pa.msu.edu and beers@pa.msu.edu}
\altaffiltext{3}{Instituto de Astrof\'{\i}sica de Canarias, 38205 La
Laguna, Tenerife, Spain; callende@iac.es}
\altaffiltext{4}{Departamento de Astrof\'{\i}sica, Universidad de La
Laguna, 38206 La Laguna, Tenerife, Spain}
\altaffiltext{5}{NSF Graduate Research Fellow}
\altaffiltext{6}{University of California Observatories}
\altaffiltext{7}{Packard Fellow}

\begin{abstract}

\noindent
We determine the average metallicities of the elements of cold halo
substructure (ECHOS) that we previously identified in the inner halo
of the Milky Way within 17.5 kpc of the Sun.  As a population,
we find that stars kinematically associated with ECHOS are chemically
distinct from the background kinematically smooth inner halo stellar
population along the same Sloan Extension for Galactic Understanding
and Exploration (SEGUE) line of sight.  ECHOS are systematically more
iron-rich, but less $\alpha$-enhanced than the kinematically-smooth
component of the inner halo.  ECHOS are also chemically distinct from
other Milky Way components: more iron-poor than typical thick-disk stars
and both more iron-poor and $\alpha$-enhanced than typical thin-disk
stars.  In addition, the radial velocity dispersion distribution of
ECHOS extends beyond $\sigma \sim 20$ km s$^{-1}$.  Globular clusters are
unlikely ECHOS progenitors, as ECHOS have large velocity dispersions and
are found in a region of the Galaxy in which iron-rich globular clusters
are very rare.  Likewise, the chemical composition of stars in ECHOS do
not match predictions for stars formed in the Milky Way and subsequently
scattered into the inner halo.  Dwarf spheroidal (dSph) galaxies are
possible ECHOS progenitors, and if ECHOS are formed through the tidal
disruption of one or more dSph galaxies, the typical ECHOS [Fe/H] $\sim
-1.0$ and radial velocity dispersion $\sigma \sim 20$ km s$^{-1}$ implies
a dSph with $M_{\mathrm{tot}} \gtrsim 10^{9}~M_{\odot}$.  Our observations
confirm the predictions of theoretical models of Milky Way halo formation
that suggest that prominent substructures are likely to be metal-rich,
and our result implies that the most likely metallicity for a recently
accreted star currently in the inner halo is [Fe/H] $\sim -1.0$.

\end{abstract}


\section{Introduction}

\defcitealias{schl09}{S09}

In $\Lambda$CDM cosmology, galaxies like the Milky Way are formed
through the stochastic accretion of smaller dark matter halos
and the continuous accretion of gas \citep[e.g.,][]{pre74,whi78}.
The statistical properties of the stochastic accretion have been
well characterized by large cosmological dark matter only $n$-body
simulations \citep[e.g.,][]{spr05}.  On the other hand, no model has
yet self-consistently incorporated the gas dynamics, star formation,
and feedback necessary to satisfactorily reproduce the formation of
Milky Way analogs.  In any case, an understanding of the statistical
properties of the accretion histories of Milky Way analogs does not
predict the unique formation history of the Milky Way.  Fortunately,
the dynamical and chemical properties of the Milky Way's thin disk,
thick disk, bulge, and halo provide strong constraints on the formation
of our own Galaxy \citep[e.g.,][]{fre02,hel08}.

The kinematics of the stellar halo of the Milky Way are simpler to
describe than the kinematics of other Galactic components.  As a
result, the stellar halo is the Galactic component in which it is
easiest to unambiguously identify the remnants of the smaller halos
that the Milky Way has accreted through its history.  For that reason,
the substructures recently discovered in the halo provide a direct
measure of accretion in the region of the Galaxy that is far easier
to model than the disk or the bulge.  The existence of substructure
in the halo of the Milky Way is now well established by star counts
\citep[e.g.,][]{tot98,tot00,ive00,yan00,ode01,viv01,gil02,new02,roc02,maj03,
yan03,roc04,duf06,bel06,gri06a,gri06b,viv06,bel07,bell08,jur08,gri09,wat09},
kinematic measurements
\citep[e.g.,][]{maj96,chi98,hel99,chi00,kep07,ive08,kle08,sea08,kle09,smi09,sta09},
and chemical abundances \citep[e.g.,][]{ive08,an09}; these observed
substructures are likely the remains of the stellar populations that
formed as part of independent, bound structures that are now being
disrupted and dispersed in the halo.

In the first paper in this series, \citet{schl09}--\citetalias{schl09}
hereafter--we identified elements of cold halo substructure (ECHOS) as
overdensities in the spatial and radial velocity distribution of the inner
halo's stellar population.  Except for those ECHOS identified along lines
of sight known to host surface brightness substructure, none of our ECHOS
are detected as overdensities of stars in surface brightness substructure.
As a result, they are distinct from surface brightness substructure
and are likely the debris of more ancient accretion events.  We used our
detections to infer that as much as 10\% of the inner halo by volume might
have 30\% of its stars in ECHOS, and we used that measurement together
with similar measurements for surface brightness substructure to suggest
that the Milky Way's accretion history has been roughly constant over the
last few Gyr with no massive ($M_{\mathrm{tot}} \sim 10^{10}~M_{\odot}$)
single accretion events in that interval.  Moreover, the distribution
of ECHOS in Galactic coordinates is consistent with isotropy given the
completeness limits of our search, possibly indicating an accretion
origin for ECHOS (rather than an association with the Galactic disk).

The chemical composition of the Milky Way's stellar populations also
illuminate their origin \citep[e.g.,][]{mcw97}.  In general, since iron is
introduced into the interstellar medium in supernovae (SN) explosions, the
[Fe/H] of a stellar population is correlated with the total integrated
star formation that occurred in that population.  Qualitatively then,
low [Fe/H] indicates relatively little integrated star formation,
while high [Fe/H] indicates relatively more integrated star formation.
If the initial mass function (IMF) of stellar populations is not a
strong function of environment or metallicity at [Fe/H] $\gtrsim -3.0$
\citep[e.g.,][]{bas10}, then a stellar population's abundances of the
$\alpha$-elements (O, Mg, Si, Ca, and Ti) relative to iron are related
to the duration of the star formation that produced that population.
Core-collapse (i.e., Type II and Type Ibc) SN result from the explosions
of stars initially more massive than about $8~M_{\odot}$ and produce
large amounts of the $\alpha$-elements relative to iron.  Since stars
that massive spend very little time on the main sequence, the enrichment
of the interstellar medium (ISM) by the ejecta from core-collapse SN
occurs very quickly, within a few Myr.  In contrast, thermonuclear
(i.e., Type Ia) SN result from the explosion of white dwarfs as a
result of runaway nuclear burning and produce large amounts of iron
relative to the $\alpha$-elements.  The exact progenitor and therefore
the characteristic timescale on which thermonuclear SN start to enrich
the ISM is still debated \citep[e.g.,][]{sca05}; nevertheless, the
timescale is longer than the comparable time to ISM enrichment through
core-collapse SN.  For those reasons, a short burst of star formation
will leave behind a stellar population with [$\alpha$/Fe] $\gtrsim 0$,
while a more extended star formation history will produce a population
with [$\alpha$/Fe] $\sim 0$ (neglecting any differences in the IMF).
Accordingly, a lower star formation rate will allow thermonuclear
SN to reduce [$\alpha$/Fe] at relatively low [Fe/H], whereas a high
star formation rate will produce many generations of stars before
thermonuclear SN begin to reduce [$\alpha$/Fe], pushing the ``knee"
in the [Fe/H]--[$\alpha$/Fe] plane to higher [Fe/H].

Stars in the inner halo typically have [Fe/H] $\sim -1.6$
and are enhanced in the $\alpha$-elements relative to iron
\citep[e.g.,][]{rya91a,rya91b,mcw95,all06}.  \citet{rob05} explained the
chemistry of most of the stellar mass in the inner halo in the context of
the $\Lambda$CDM model of galaxy formation with the accretion of a massive
$M_{\mathrm{tot}} \sim 5 \times 10^{10}~M_{\odot}$ halo $\sim 10$ Gyr in
the past.  The high-mass and short-timescale for star formation in such
a massive progenitor of the inner halo are consistent with the observed
chemistry.  This scenario is in contrast to the composition of the
classical dwarf spheroidal galaxies (dSph), which at the average [Fe/H] of
the inner halo have [$\alpha$/Fe] closer to solar \citep[e.g.,][]{mat98}.
At the same time, the \citet{rob05} model did not directly address the
origin of the substructure now known to exist in the inner halo, and
those substructures are likely related to more recent accretion events.

In the same way, the chemical properties of substructure reveal
something about its origin.  Since ECHOS are likely the remains
of recent accretion events ($\tau \lesssim 5$ Gyr), their chemical
composition constrains the properties of a more recently-disrupted
class of contributors to the stellar population of the inner halo than
the few massive progenitors that contributed the bulk of the inner
halo's stellar population.  If ECHOS are the remains of disrupted
dSph galaxies, then the dSph luminosity--metallicity relation can
be used to infer the luminosity of a progenitor from its mean iron
metallicity \citep[e.g.,][]{hel06b,kir08b}.  Indeed, the \citet{kir08b}
luminosity--metallicity relation for dSph galaxies is now well calibrated
over four decades in luminosity using.  Combined with stellar mass to
light ratios, the \citet{kir08b} luminosity--metallicity relation can
be turned into a stellar mass--metallicity relation.  As a result, the
average iron metallicity of a dSph stellar population can be considered a
proxy for its stellar mass.  Likewise, the [$\alpha$/Fe] of substructure
indicates the duration of star formation in its progenitor.  Typical halo
stars have [$\alpha$/Fe] $\gtrsim 0.3$ at [Fe/H] $\sim -1.6$ indicating
formation in a truncated episode of star formation.  In contrast, stars
in classical dSph typical have [$\alpha$/Fe] $\sim 0$ at [Fe/H] $\sim
-1.6$, indicating formation in an episode of prolonged star formation.

We measure the chemical properties of the ECHOS discovered in
\citetalias{schl09} and use their combined kinematic and dynamical
properties to better understand their origin in the context of the
$\Lambda$CDM model of Galaxy formation.  This paper is organized as
follows: in Section 2 we describe the data we use in this analysis.
In Section 3 we describe how we derive average [Fe/H] and [$\alpha$/Fe]
for each ECHOS through the analysis of coadded metal-poor main sequence
turnoff star spectra creating using spectra from individual metal-poor
main sequence turnoff stars kinematically associated with each ECHOS.
In Section 4 we discuss the implications of our findings for the formation
of the Milky Way.  We summarize our conclusions in Section 5.

\section{Data}

The Sloan Extension for Galactic Understanding and Exploration
(SEGUE) survey observed approximately 240,000 Milky Way stars with
apparent magnitudes in the range $14 < g < 20.3$ with the fiber-fed
Sloan Digital Sky Survey (SDSS) spectrograph at moderate resolution.
Spectroscopic targets were selected from the combined 11,663 deg$^2$
$ugriz$ photometric footprint of the SDSS and SEGUE.  The SDSS telescope
and spectrograph obtain $R \approx 1800$ spectra between 3900 \AA~and 9000
\AA~with high spectrophotometric accuracy.  The SEGUE instrumentation,
data processing pipelines, survey strategy, along with radial velocity
and atmospheric parameter accuracies are described in \citet{yan09},
\citet{lee08a,lee08b}, \citet{all08}, and the SDSS-II DR7 paper
\citep{abaz09}.  The SDSS survey is described in detail in \citet{fuk96},
\citet{gun98}, \citet{yor00}, \citet{hog01}, \citet{smi02}, \citet{pie03},
\citet{ive04}, \citet{gun06}, \citet{tuc06}, and \citet{pad08}.

In \citetalias{schl09} we examined a sample of 10,739 metal-poor main
sequence turnoff stars collected from 137 seven deg$^2$ lines of sight.
These metal-poor main sequence turnoff (MPMSTO) stars have both the
$g-r$ color and significant UV excess expected for the main sequence
turnoff of a metal-poor population (for a detailed description of
the MPMSTO sample see Section 2 of \citetalias{schl09}).  Given the
magnitude limits of the SEGUE survey, the MPMSTO sample was selected
because MPMSTO stars are the highest-density tracer of the inner halo.
In \citetalias{schl09}, we defined the inner halo as the volume more
than 10 kpc from the Galactic center, within 17.5 kpc of the Sun, and
more than 4 kpc from the Galactic plane.  Though we found no reason
to reject a kinematically smooth model for the inner halo on average,
we discovered many radial velocity overdensities that we termed ECHOS.
We identified ECHOS along individual lines of sight along which there
was a very significant deviation from a kinematically smooth model of the
inner halo.  We gave these substructure the name ECHOS to differentiate
them from surface brightness substructure like tidal streams, as the
only ECHOS we could relate to surface brightness substructures were
those ECHOS discovered along lines of sight targeted at known surface
brightness substructures.  ECHOS are also distinct in that they are
likely related to more ancient accretion events than surface brightness
substructure in the same volume \citep[e.g.,][]{joh08}.  As the same
time, ECHOS are likely related to more recent accretion events than
those substructures that will be discovered in the same volume using
the 6D kinematic data that will become available from Gaia and the
LSST \citep[e.g.,][]{hel06a,mcm08,gom10}.  The ECHOS we discovered were
preferentially bunched at the faintest apparent magnitudes of the SEGUE
spectroscopic sample; indeed the 25\%, 50\%, and 75\% percentile in
$r$-magnitude and spectral signal to noise per 1 \AA~pixel (abbreviated
S/N hereafter) were (19.0,19.5,19.8) and (7.6,10.4,15.4) respectively.
As a result, the spectra of individual MPMSTO stars were generally at
too low a S/N to precisely measure abundances.

\section{Analysis}

We coadd all stellar spectra kinematically associated with an ECHOS in a
narrow range of effective temperature and analyze the resultant coadded
spectrum to determine the average abundance of the ECHOS.  At the same
time, we use an equivalent coaddition process to determine the average
metallicity of the MPMSTO stars in the kinematically smooth inner halo
population along the same line of sight.

We use these measurements to compare the mean chemical abundance
of the ECHOS and smooth halo population.  Both calculations are subject
to the same systematics, so any apparent chemical offset can only
result from an underlying chemical difference or from random effects.
To quantify the effects of randomness, we compute the precision and
accuracy of our technique by analyzing two classes of objects with known
composition.  The first class of objects are individual MPMSTO stars
that have been observed both at high S/N during the SEGUE survey and
at high resolution by larger telescopes.  The high S/N SEGUE spectra
can be degraded to arbitrarily low S/N by a detailed noise model so
that we can test the coaddition process with data representative of
the low S/N spectra available for MPMSTO stars in ECHOS.  The second
class of objects are MPMSTO stars associated with the globular clusters
M~13 and M~15.  Using those two test cases, we quantify the mean square
error (MSE $\equiv$ bias$^2$ + variance) of our metallicity estimates.
We describe these steps in detail in the following subsections.

\subsection{The SEGUE Stellar Parameter Pipeline}

The SEGUE Stellar Parameter Pipeline \citep[SSPP -
][]{lee08a,lee08b,all08} uses Sloan spectroscopy and $ugriz$ photometry to
infer the stellar atmosphere parameters ($T_{\mathrm{eff}}$, $\log{g}$,
[Fe/H], and [$\alpha$/Fe]) of stars observed in the course of SDSS
and SEGUE.  The SSPP implements a multimethod algorithm in which
many different techniques are used to compute the stellar parameters.
The SSPP then averages the result of each method known to be valid in
a given color and S/N range to determine the final $T_{\mathrm{eff}}$,
$\log{g}$, [Fe/H], and [$\alpha$/Fe] reported for all stars observed in
the SDSS and SEGUE surveys.

\citet{lee08a,lee08b} determined the accuracy and precision of the
SSPP in three ways.  First, they compared the atmospheric parameters
determined by the SSPP from high S/N SEGUE spectra with the atmospheric
parameters determined from high-resolution spectroscopy from HIRES and
ESI on the Keck Telescopes, HRS on the Hobby-Eberly Telescope (HET), and
HDS on the Subaru Telescope.  Their results suggest that including both
systematic and random error, the SSPP has a one-sigma precision of 141
K in $T_{\mathrm{eff}}$, 0.23 dex in $\log{g}$, and 0.23 dex in [Fe/H].
The stars bright enough to be observed at high resolution all had S/N
$\gtrsim 50$ SEGUE spectra, so direct comparison with high-resolution
spectra can only be made for high S/N spectra.  For that reason, they
degraded these high S/N SEGUE spectra with stellar parameters well
characterized by high-resolution spectroscopy with a detailed noise model
to create many thousand spectra at many values of S/N between 55 and 1.
They then ran the SSPP on the noise-degraded spectra to determine the
accuracy and precision of the SSPP as a function of S/N and reported the
result in in Table 6 of \citet{lee08a}.  They found a [Fe/H] precision of
(0.5,0.2,0.1) dex at S/N of (5,10,15) and a [$\alpha$/Fe] precision of
0.1 dex at S/N greater than 20 \citep{lee11}.  Finally, \citet{lee08b}
applied the SSPP to stars associated with open and globular clusters,
where they found that the SSPP achieves a precision in [Fe/H] of 0.13
dex over the range $-0.3 \leq g-r \leq 1.3$, $2.0 \leq \log{g} \leq 5.0$,
and $-2.3 \leq$ [Fe/H] $\leq 0.0$.

\subsection{Coaddition Algorithm}

We include in the coadded spectra only those spectra that correspond
to MPMSTO stars with radial velocities that indicate membership in the
population of interest.  In addition, we include in the coadded spectra
only those spectra that correspond to MPMSTO stars in a finite range
of effective temperature, as the strength of spectral lines is affected
by temperature as well as [Fe/H].  Consequently, the $T_{\mathrm{eff}}$
range has to be small enough to ensure that the coadd gives an accurate
[Fe/H] that is representative of the population.  In order to estimate
the effective temperature of a MPMSTO star from its $g-r$ color, we fit a
linear model between $g-r$ color and $T_{\mathrm{eff}}$ in the range $0.15
< g-r < 0.5$ for all stars in SDSS DR7 with SEGUE spectroscopy at S/N $>
40$ and a SSPP [Fe/H] $< -1.0$.  We find that $T_{\mathrm{eff}} \approx
-3800 (g-r) + 7300$ with about 500 K of scatter at constant $g-r$ color.
We use $g-r$ color because the photometric accuracy of the SDSS does
not vary over the apparent magnitude range of our MPMSTO sample, while
the spectral S/N and therefore $T_{\mathrm{eff}}$ precision inferred
from the spectra varies substantially.  Moreover, there are no reliable
$T_{\mathrm{eff}}$ estimates for the half our MPMSTO sample with S/N $<
10$.  For those reasons, we use the simple relationship between $g-r$
color and $T_{\mathrm{eff}}$ at high S/N to ensure that we reliably
include in each coadd spectrum only MPMSTO star spectra in a narrow
range of $T_{\mathrm{eff}}$ even at low S/N.  Likewise, the photometric
selection and spectroscopic confirmation of MPMSTO stars indicates that
the stars in our sample have similar surface gravities.  As a result, we
ensure that each spectrum that enters into a coadd correspond to a star
in a narrow range of $T_{\mathrm{eff}}$ and $\log{g}$.  Therefore, the
only unconstrained stellar parameter is metallicity, and a SSPP analysis
of the coadded MPMSTO spectrum will produce an unbiased estimate of the
mean metallicity of the MPMSTO population.

We shift each MPMSTO spectrum eligible for inclusion in a coadd to
a heliocentric radial velocity $v_r = 0$ km s$^{-1}$.  We then use
natural cubic spline interpolation to interpolate both the spectrum
and its inverse variance on to a common grid in wavelength.  Next, we
numerically integrate the area under the curve defined by the spectrum and
normalize both the spectrum (by dividing by the normalization factor) and
the inverse variance (by multiplying by the normalization factor squared)
to ensure that each spectrum that is to be included in the coadd has the
same scale.  For each population of interest, we then create an ensemble
of realizations of the coadded spectrum by bootstrap resampling from the
set of radial velocity zeroed, interpolated, and normalized spectra that
belong to that population.  Each spectrum contributes to each wavelength
bin in proportion to its inverse variance in that bin relative
to the other spectra selected for coaddition.  One danger to this
approach is the possibility that the resultant coadded spectrum does not
correspond to the spectrum of any physical star.  This is unlikely in
our analysis though, as we obtain good agreement between globular cluster
metallicities produced by coaddition and their known metallicities from
high resolution spectroscopy.  We describe the coaddition process in
detail in Appendix A.

The SSPP also uses Sloan $ugriz$ photometry in its parameter estimation
routines, so we determine an equivalent $ugriz$ photometric measurement
for our coadd spectra by computing a weighted average of the $ugriz$
photometric measurement of the individual stars in each bootstrap coadd,
using the mean S/N between 3950 \AA~and 6000 \AA~as the weight.  We then
run the SSPP on each of the bootstrap coadded spectra and average $ugriz$
photometry to determine the mean of the SSPP [Fe/H] and [$\alpha$/Fe]
estimates for that particular MPMSTO population.

\subsection{Accuracy and Precision in the Ideal Case}

To determine the precision and accuracy of our coaddition algorithm as a
function of S/N and population metallicity, we analyzed spectra created
by coadding individual MPMSTO star spectra that had been observed to
very high S/N during the SEGUE survey but that had been subsequently
degraded with a detailed noise model.  Though this is the ideal case
of uniform effective temperature, surface gravity, and metallicity, the
result will reveal the amount of bias and variance in our measurements
that can be attributed to noise and population metallicity.

Section 6 of \citet{lee08a} describes the algorithm used to create the
noise degraded spectra we use in this analysis and summarize here.
Each SEGUE plate obtains spectra for stars that span a range of
magnitudes from $g \approx 15$ to $g \approx 20$, but is exposed to
a fiducial S/N for the faint targets.  As such, the noise properties
of the SEGUE spectra vary with the magnitude and color of the targets,
with increasing fractional contribution from the sky for faint targets.
Because the observing criteria for the survey were homogeneous, it is
possible to parametrize those variations and create a realistic model
noise spectrum that can be used to create low S/N realizations of high
S/N spectra.  Randomly chosen residual spectra from the sky fibers are
then added to complete each realization.  This noise model was used in
\citet{lee08b} to test the accuracy of the SSPP, and we use the same
model here to test the accuracy of the SSPP parameters of our coadds.
We use a sample of 640 noise-added realizations of eight high S/N SEGUE
MPMSTO stars with [Fe/H] values from $-2.41 <$ [Fe/H] $< -0.23$, which
spans the metallicity range of our sample.  We have 54 realizations at
S/N values from 7.5 to 55.

We use this large sample of noise-degraded spectra for eight MPMSTO stars
to determine the precision and accuracy of our coaddition algorithm
as a function of metallicity and S/N.  The analysis of noise-added
spectra allows us to examine the performance of the SSPP and our
coaddition algorithm over a range of metallicity and S/N that spans
our entire sample.  We plot the results of our analyses for [Fe/H] in
Figure~\ref{fig01} and for [$\alpha$/Fe] in Figure~\ref{fig02}.  The MSE
of our SSPP analysis of coadded spectra created from the coaddition of
noise-degraded spectra of single MPMSTO stars ranges from 0.05 dex in
both [Fe/H] and [$\alpha$/Fe] for the most iron-rich stars to 0.20 dex
in both [Fe/H] and [$\alpha$/Fe] for the most iron-poor stars.  In both
cases, the range in effective S/N results because the coadded spectra
are created by coadding from an ensemble of spectra with different S/N,
and the range of S/N apparent in Figure~\ref{fig01} and Figure~\ref{fig02}
is larger than the equivalent range of S/N in the ECHOS.  Reassuringly,
there is no obvious trend in the precision and accuracy of the bootstrap
coaddition process with effective temperature.  The outliers at low
S/N in Figure~\ref{fig01} and Figure~\ref{fig02} are likely coadded
spectra that include very low S/N spectra with large sky residuals,
possibly taken from SEGUE plates with below-average sky subtraction.
The existence of outliers due to this effect is another reason why the
bootstrap resampling from many spectra is important to determine the
error distribution that results from the coaddition process.  In the next
subsection, we will use this analysis of the noise-added data to determine
the MSE of our coadditon analysis as function of metallicity and S/N.

\subsection{Accuracy and Precision in a Representative Case}

In reality, the spectra in each ECHOS coadd correspond to stars in a
small but finite range of effective temperature, surface gravity, and
composition.  In order to determine the degree to which these spreads
in properties affects our ability measure the mean metallicity of the
ECHOS and smooth halo populations by coadding spectra belonging to
each, we analyzed spectra created by coadding individual MPMSTO spectra
corresponding to MPMSTO stars that likely belong to the well studied
Galactic globular clusters M~13 and M~15.  The M~13 and M~15 data are
representative of the range in $T_{\mathrm{eff}}$ and $\log{g}$ of a real
MPMSTO population as selected by the SEGUE survey.  It is exactly the fact
that the globular cluster MPMSTO stars occupy a narrow range in [Fe/H]
that makes the this globular cluster data so useful--we know the expected
[Fe/H] very well and can test whether or not we converge to the known
value when coadding from a range of $T_{\mathrm{eff}}$ and $\log{g}$.
Unfortunately, [Fe/H] and $T_{\mathrm{eff}}$ are degenerate, so we
optimize the $g-r$ range selected in order to produce coadded spectra
with the maximum S/N and therefore the most accurate abundance estimate
from the SSPP in the following way.  Increasing the $T_{\mathrm{eff}}$
range of spectra that are coadded increases the number of spectra
included in each coadd, increases the total signal in each coadd, and
ultimately increases the precision of the abundance estimate.  The trade
off is that coadding spectra in too large a range in $T_{\mathrm{eff}}$
can decrease the accuracy of the abundance estimate.  We find that
coadding spectra that belong to stars in 250 K bins produces the same
accuracy as the coaddition of stars in 500 K bins, as the scatter in the
$T_{\mathrm{eff}}$--$g-r$ relation at constant $g-r$ color is about 500 K.
Combined with the fact the coadd spectrum that results from coadding
spectra in the 500 K bin always reaches higher S/N (and therefore higher
precision) than the coadd spectrum that results from coadding spectra in
the 250 K bin, we exclusively use the 500 K bin in our analysis of ECHOS.

We select those stars that have equatorial coordinates that place them
within the tidal radius of each cluster \citep[as reported in][]{har96}
and that have radial velocities consistent with cluster membership.
Given the precision of SEGUE radial velocities at the S/N of the cluster
spectra for cluster members that meet the SEGUE turnoff sample criteria,
that corresponds to 15 km s$^{-1}$ for M~13 and 25 km s$^{-1}$ for M~15.
We select for coaddition only those spectra that correspond to MPMSTO
stars with $g-r$ colors that place them within the 500 K bin in effective
temperature that produces the highest S/N coadd.  The median S/N of
the MPMSTO spectra that remain after applying these cuts is 22.9 from
12 spectra for M~13 and 9.0 from 11 spectra for M~15.  The number of
spectra in each globular cluster coadd is approximately equivalent to
the number of spectra included in each ECHOS coadd.

From the $n$ MPMSTO spectra that remain after the application of our cuts
in equatorial coordinate, radial velocity, and $g-r$ color, we select
with replacement $m=1,2,3,\ldots,n$ spectra from the available data.
We coadd the spectra and use the SSPP to derive [Fe/H] and [$\alpha$/Fe]
for that bootstrap coadded spectrum, and save the result.  We repeat
that process 100 times for each of $m=1,2,3,\ldots,n$ to build up
the distribution of SSPP [Fe/H] and [$\alpha$/Fe] estimates for both
globular clusters as a function of S/N.  For M~13, the SSPP produces an
estimate of [Fe/H] $= -1.7 \pm 0.15$ and [$\alpha$/Fe] $= 0.3 \pm 0.15$.
High-resolution measurements of M~13 by \citet{kra97} and \citet{coh05}
yielded [Fe/H] $= -1.59$ with [$\alpha$/Fe] $=0.22$ and [Fe/H] $=
-1.55$ with [$\alpha$/Fe] $=0.26$ respectively.  Likewise, for M~15 we
find that [Fe/H] $= -2.4 \pm 0.2$ and [$\alpha$/Fe] $= 0.3 \pm 0.2$.
High-resolution measurements of M~15 by \citet{sne97} and \citet{sne00b}
yielded [Fe/H] $= -2.19$ with [$\alpha$/Fe] $=0.38$ and [Fe/H] $= -2.28$
with [$\alpha$/Fe] $=0.40$ respectively.  We present the results for M~13
in Figure~\ref{fig03} and the results for M~15 in Figure~\ref{fig04}.
This implies that at the resolution of the SEGUE spectra and for the
range of $T_{\mathrm{eff}}$ and $\log{g}$ included in the coadds,
S/N is the dominant contribution to the MSE in estimating the average
metallicity of a population using the ensemble of bootstrap coadds.
At the same time, these results compare favorably to those presented in
\citet{lee08a,lee08b} and \citet{all08}, as we measure the performance of
the SSPP on a subset of the SEGUE data in a narrow range in $g-r$ color.

We summarize the precision and accuracy of our SSPP analysis of coadded
spectra in Figure~\ref{fig05}.  The precision and accuracy of our SSPP
analysis is a function of metallicity; consequently, our estimates are
most precise and accurate ($\sim 0.1$ dex in [Fe/H] and [$\alpha$/Fe]) for
the most metal-rich populations, and least precise and accurate for the
most metal-poor populations ($\sim 0.2$ dex in [Fe/H] and [$\alpha$/Fe]).
Moreover, we find that at equivalent metallicity, the MSE we compute for
M~13 and M~15 based on coadd spectra created by coadding MPMSTO spectra in
a finite range of effective temperature and surface gravity are in good
agreement with the MSE computed in the ideal case of constant effective
temperature and surface gravity for coadds based on noise-degraded
MPMSTO spectra.  As a result, we use the precision we derived for the
SSPP metallicity analysis of the noise-added spectra to characterize
the precision of our ECHOS metallicity measurements.  For that reason,
including both statistical and systematic effects, our SSPP analysis of
coadded SEGUE MPMSTO spectra produces estimates that are sufficiently
precise and accurate to identify chemical differences between ECHOS and
the kinematically smooth inner halo MPMSTO population on the order $\sim
0.2$ dex in both [Fe/H] and [$\alpha$/Fe].  We use the expected MSE as a
function of metallicity and S/N given in Figure~\ref{fig05} to determine
whether an ECHOS is chemically distinct from the kinematically smooth
halo MPMSTO population along the same line of sight.

\subsection{The Metallicity of ECHOS}

For each of the three classes of ECHOS from \citetalias{schl09},
we consider for coaddition those spectra that correspond to stars
within a radial velocity overdensity and therefore consistent with
ECHOS membership.  As defined in \citetalias{schl09}, a radial velocity
overdensity is a group of MPMSTO stars observed with radial velocities
within a narrow range such that the group is extraordinarily unlikely
to be observed if the underlying population were kinematically smooth.
For the bin detections, we coadd the spectra of stars in the 20 km
s$^{-1}$ bin that contains the significant detection.  For the peak
detections, we coadd the spectra of stars within the measured width of
the peak in the cumulative distribution function (the $\Theta$ statistic
of \citetalias{schl09}).  The width of the peak in $\Theta$ is the same
as the velocity dispersion quoted for each ECHOS in Tables~\ref{tbl-2}
and~\ref{tbl-3}.

As before, we select those stars within the 500 K wide bin in effective
temperature that produces the highest S/N in the resultant coadded
spectrum.  We shift each candidate spectrum to $v_r = 0$ km s$^{-1}$
and interpolate the spectrum and its inverse variance on to a common
wavelength grid, and rescale each spectrum and its inverse variance to
a common scale.  From the $n$ radial velocity selected, temperature
controlled, radial velocity zeroed, interpolated, and scaled MPMSTO
spectra, we sample $n$ with replacement and coadd the spectra to
create a single bootstrap realization of the resultant coadded spectrum
distribution.  For each ECHOS, we repeat the resampling process 100 times
to create 100 bootstrap coadded spectra.  We then run the SSPP on all of
the bootstrap coadded spectra to determine the distribution of measured
[Fe/H] and [$\alpha$/Fe] for the ensemble of realizations.

As a control, along the same line of sight as each ECHOS we perform
the same steps on all the MPMSTO spectra that are associated with the
kinematically smooth halo population.  As a result, the stars in the
control sample are in the same volume, were observed at the same time,
and have a similar range of magnitude and S/N as the stars in the ECHOS.
In that way, we can characterize the mean of [Fe/H] and [$\alpha$/Fe]
and estimate the error in our estimate of the mean for both the ECHOS
and the kinematically smooth population along the same line of sight.
With that information, any observed difference in composition is unlikely
to result from systematic effects.  Consequently, any observed chemical
offset between an ECHOS and the kinematically smooth inner halo MPMSTO
population along the same line of sight likely results from genuine
chemical differences.

We report the results of these calculations for all of the ECHOS
discovered in \citetalias{schl09} in Figure~\ref{fig06} through
Figure~\ref{fig11} and in tabular form in Table~\ref{tbl-1} through
Table~\ref{tbl-3}.  In Figure~\ref{fig06} through Figure~\ref{fig11},
the precision of our SSPP analysis is best characterized by the scatter
apparent in results of the bootstrap coaddition process for both
ECHOS and the kinematically smooth halo.  If the two clouds of points
do not overlap, than the two populations are chemically distinct.
We quantify this scatter and report the result in Table~\ref{tbl-1}
through Table~\ref{tbl-3}.  In all cases where the quoted metallicity
of the smooth component is more iron-rich than typically associated
with the smooth inner halo, the reason is because the ECHOS dominates
the MPMSTO population along that line of sight (see Figures 2 through
11 of \citetalias{schl09}).  As a result, it is difficult to identify
a large enough sample of stars in the kinematically smooth inner halo
MPMSTO population for the equivalent analysis, so the quoted metallicity
of the smooth halo can be significantly influenced by the metallicity
of the ECHOS.

\section{Discussion}

\subsection{The Kinematic and Chemical Properties of ECHOS}

We searched for correlations between the dynamical properties of ECHOS
from \citetalias{schl09} and the chemical properties determined in
this analysis.  For the rest of this section, we examine the properties
of the 21 Class II peak ECHOS from Table 3 of \citetalias{schl09}.
That sample of 21 ECHOS is both the largest sample and the most
representative sample of the inner halo ECHOS population.  In general,
we find that ECHOS that are iron-rich also have large radial velocity
dispersions and are the most prominent ECHOS in that they have high
number densities and are fractionally the largest contributors to the
MPMSTO population along the line of sight where they were discovered;
we plot these relations in Figure~\ref{fig12}.  Though large velocity
dispersions are found only for the most metal-rich ECHOS, small velocity
dispersions are found at all metallicities.  The prominence of metal-rich
substructures was predicted by \citet{fon06}, as metal-rich substructures
are preferentially produced by the most luminous progenitors.  We find
no significant correlation between [Fe/H] or [$\alpha$/Fe] and distance.

We initially identified ECHOS because they are kinematically distinct
from the kinematically smooth inner halo MPMSTO population.  As we showed
in Figure~\ref{fig06} through Figure~\ref{fig11}, nearly all ECHOS are
also chemically distinct from the background smooth inner halo MPMSTO
population along the same line of sight.  We summarize this chemical
distinctiveness in Figure~\ref{fig13}.  As a population, ECHOS are
more iron-rich but less $\alpha$-enhanced than the kinematically smooth
background inner halo MPMSTO population.  We showed in \citetalias{schl09}
that 10\% of the inner halo (by volume) has 30\% of its MPMSTO population
in ECHOS.  Combined with the observation that ECHOS are metal-rich,
these facts suggest that the most likely metallicity for an accreted
star in the inner halo is [Fe/H] $\sim -1$.  At ECHOS [Fe/H] $\gtrsim
-1$, the apparent correlation between the iron metallicity of ECHOS
and the iron metallicity of the smooth component of the halo results
from the fact that iron-rich ECHOS are also the most prominent ECHOS
(see Figure~\ref{fig12}).  That is, since the iron-rich ECHOS make up
such a large fraction of the halo MPMSTO population along the line of
sight where they were discovered, it is difficult to isolate a sample
of MPMSTO stars in the kinematically smooth halo population.

To asses the statistical significance of the observation that the
population of ECHOS is more iron-rich but less $\alpha$-enhanced than the
kinematically smooth background inner halo MPMSTO population, we used a
Monte Carlo simulation.  Imagine that ECHOS and the kinematically smooth
component of the halo really did have the same chemical composition.
In that case, in Figure~\ref{fig13} the departure of the points from
the line $y=x$ must result from imperfect observation, characterized by
the error bars in the plot.  Under that null hypothesis, we sample each
ECHOS composition from a normal distribution centered on the line $y=x$
with standard deviation equal to the error in our measurement of the
composition of the ECHOS.  Likewise, we sample the composition of the
smooth component along each line of sight from a normal distribution
centered on the same point on the line $y=x$ with standard deviation
equal to the error in our measurement of the composition of the smooth
component.  We compute the signed, cumulative Euclidean distance of the
entire population from the line $y=x$, and save the result.  We repeat
this process 10$^6$ times.  We find that under the null hypothesis,
in no instance does the Monte Carlo simulation produce a cumulative
distance of each point from the line $y=x$ equal to the cumulative
distance observed in the ECHOS [Fe/H] distribution.  Therefore the
probability that the population of ECHOS has the same [Fe/H] as the
smooth population along each line of sight in which we discovered an
ECHOS is less than 10$^{-6}$.  Our [$\alpha$/Fe] estimates are much
less precise than our [Fe/H] estimates; nevertheless, the probability
that the population of ECHOS has the same [$\alpha$/Fe] as the smooth
population along each line of sight in which we discovered an ECHOS
is about 10$^{-3}$.  The fact that ECHOS, as a population,
are so chemically distinct from the smooth background inner halo MPMSTO
population along the same line of sight strongly supports the kinematic
substructure identifications in \citetalias{schl09}.

The stars in ECHOS preferentially have apparent magnitudes that
place them in the most distant half of our MPMSTO sample \citep{schl09}.
As a result, it is possible that the chemical distinctiveness of ECHOS
is the result of a metallicity gradient in the inner halo.  To test this
hypothesis, we considered only those lines of sight that were not targeted
at a known element of substructure and for which we had no significant
ECHOS detection of any kind.  In other words, those lines of sight which
are dominated by the smooth component of the halo.  For every MPMSTO
star in that sample, we very roughly estimate the distance to each star
according to the procedure described in Section 2 of \citetalias{schl09}.
We then split the sample in half at about 14 kpc: the nearest 50\% of the
MPMSTO population goes into the ``close" subsample and the farthest 50\%
of the MPMSTO population goes into the ``far" subsample.

Then, for each subsample, we compute line of sight average
metallicities by the same spectral coaddition process described in
Section 3.  In that way, we end up with two estimates for the average
chemical composition of MPMSTO stars in the smooth component along each
line of sight where there is no significant substructure -- an estimate
for the ``close" subsample and estimate for the ``far" subsample.  We find
that the average metallicity for the ``close" subsample is [Fe/H] $=
-1.7 \pm 0.1$ while the average metallicity for the ``far" subsample
is [Fe/H] = $-1.5 \pm 0.1$.  Meanwhile, the average metallicity of our
ECHOS is [Fe/H] = $-1.1 \pm 0.1$.  For that reason, the apparent chemical
distinctiveness of ECHOS is unlikely to be the result of a metallicity
gradient in the inner halo.  Indeed, though the halo is potentially
chemically inhomogeneous on large scales \citep[e.g.,][]{caro07,caro10},
our analysis is confined to a relatively small region in the inner halo.

ECHOS are also chemically distinct from the other components of the Milky
Way.  The average level of star formation appears to have been more or
less continuous in the thin disk for many Gyr \citep[][]{roc00a,roc00b},
so thermonuclear SN have had plenty of time to contribute to the chemistry
of the ISM.  Consequently stars in the thin disk are typically close
to solar in both [Fe/H] and [$\alpha$/Fe].  The origin of thick disk
stars is unclear, though they are uniformly more $\alpha$-enhanced than
thin-disk stars at constant [Fe/H] \citep[e.g.,][]{ben05,red06,ben07}.
In Figure~\ref{fig14} we plot our ECHOS in the [Fe/H]--[$\alpha$/Fe]
plane along with [Fe/H] and [$\alpha$/Fe] estimates for individual
stars from \citet{edv93}, \citet{nis97}, \citet{han98}, \citet{ful00},
\citet{pro00}, \citet{ste02}, \citet{ben03}, and \citet{red03} and
presented in \citet{ven04}.  We plot only those stars from \citet{ven04}
that have a better than 50\% assocation probability with the thin disk,
thick disk, or halo as indicated by their $U$, $V$, and $W$ velocities.
ECHOS are found in a region of the [Fe/H]--[$\alpha$/Fe] plane sparsely
populated--but not completely devoid of--individual stars.  In general,
ECHOS are: (1) more iron-rich and less $\alpha$-enhanced than halo stars,
(2) more iron-poor than typical thick-disk stars, and (3) more iron-poor
and more $\alpha$-enhanced than typical thin-disk stars.  As a result,
most stars associated with recent accretion events are more metal-rich
than the average metallicity of the inner halo.  In Figure~\ref{fig15}, we
plot the location of ECHOS in the [Fe/H]--[$\alpha$/Fe] plane along with
the location of individual giant stars in eight dSph galaxies reported
in \citet{kir10}.  If ECHOS are the debris of a tidally disrupted dSph,
Figure~\ref{fig15} indicates that the progenitor may have been comparable
to the Sculptor or Leo I dSph galaxies.

Interestingly, there may be a hint in the \citet{ven04} compilation
that those stars that are on retrograde orbits are the halo stars
with [$\alpha$/Fe] most like the stars in the ECHOS (though the
ECHOS have higher iron metallicity).  Several studies using nearby
halo star samples have found correlations between increased scatter
or peculiar elemental abundance patterns and extreme kinematics
\citep{carn97,ful02,iva03,roe09,caro07,caro10}. These studies find that
metal-poor stars belonging to the distant, outer halo originated in a
more varied and/or inhomogeneous environment, in support of the idea that
the outer halo is assembled by more recent accretion of many low-mass
systems.  However, those studies are limited to [Fe/H] $< -1$, and our
results demonstrate that for stars currently in the inner halo region
of the Galaxy, most stars accreted in the last 5 Gyr are not metal-poor,
but instead have [Fe/H] $\sim -1$.  Any comprehensive study of the halo
accretion history must include stars over a broad range in metallicity.
A direct comparison between our results and the metal-poor samples
would require an analysis of selection biases in those samples that is
beyond the scope of this paper \citep[for a discussion of the issues
see][]{roe09}, but we note that the sample of halo stars in which the
ECHOS were identified was selected using ultraviolet excess, which would
tend to bias our sample against metal-rich stars (see \citetalias{schl09}
for details).

In Figure~\ref{fig16} we compare the velocity dispersions of ECHOS from
\citetalias{schl09} with the median radial velocity precision of the
MPMSTO stars in each ECHOS.  In this case, the radial velocity precision
as a function of S/N we used in \citetalias{schl09} assumed a population
metallicity of [Fe/H] $\sim -1.6$.  SEGUE radial velocity precision
becomes better at higher metallicities, so the fact that ECHOS have [Fe/H]
$\sim -1.0$ may indicate the actual radial velocity precision is somewhat
better.  In any case, we find many of our ECHOS have velocity dispersions
that are substantially larger than the expected radial velocity precision.

\subsection{The Origin of ECHOS}

The stars in ECHOS are more iron-rich and less $\alpha$-enhanced then the
typical inner halo MPMSTO star population.  They are also more iron-poor
than typical thick-disk stars but more $\alpha$-enhanced than typical
thin-disk stars.  The high [Fe/H] metallicity of ECHOS almost certainly
rules out ultrafaint dSph galaxies as ECHOS progenitors, as ultrafaint
dSph galaxies have average [Fe/H] $\lesssim -2$ \citep[e.g.,][]{kir08b}.
If ECHOS are the tidal debris of one or more dSph galaxies, the dSph
luminosity--metallicity relation \citep[e.g.,][]{kir08b} implies
a progenitor luminosity of $L \sim 10^8~L_{\odot}$ to produce a
characteristic iron metallicity [Fe/H] $\sim -1.0$.  That luminosity
combined with a reasonable dSph stellar mass to light ratio $M_{\ast}/L_V
\sim 3$ and total mass to light ratio $M_{\mathrm{tot}}/L_V \sim 10$
\citep[e.g.,][]{mat98} implies the accretion of a progenitor with
stellar mass $M_{\ast} \sim 3 \times 10^{8}~M_{\odot}$ and total mass
$M_{\mathrm{tot}} \sim 10^{9}~M_{\odot}$, comparable to Local Group
members NGC 147 or NGC 185.  Radial velocity substructures persist for
many crossing times, and because the typical crossing time in the inner
halo is 50 Myr ECHOS should be observable for up to 5 Gyr (see Section
5.1 of \citetalias{schl09}).  As a result, if the debris of an ECHOS
progenitor is still visible as cold radial velocity substructure,
the accretion event must have occurred in the last $\sim 5$ Gyr,
or equivalently, since $z = 0.5$.  Such accretion events are common
in the accretion histories of Milky Way analog halos, and a typical
halo might have experienced 10 such accretion events since $z =
0.5$ \citep[e.g.,][]{ste08,fak10}.  Again, the distribution of ECHOS
in the [Fe/H]--[$\alpha$/Fe] plane is similar to the distribution in
[Fe/H]--[$\alpha$/Fe] plane of individual giant stars in the Sculptor dSph
\citep{kir10}.

The radial velocity dispersion distribution of ECHOS may also be a clue
to their origin.  Collisionless dynamics implies that as the position
space component of the phase-space distribution of a tidally disrupted
stellar system becomes hotter, the velocity component of its phase-space
distribution must necessarily become colder \citep[e.g.,][]{bin87}.
In other words, as a tidally disrupted stellar system spreads
across the sky its velocity dispersion measured over a small patch
of sky must decrease with time.  For that reason, the radial velocity
dispersion of an element of cold halo substructure is a lower limit on
the radial velocity dispersion of its progenitor.  At the same time,
multiple wraps of stream-like substructure might give the appearance
of a single substructure with large velocity dispersion, though there
is both observational evidence \citep[e.g.,][]{bell08} and theoretical
modeling \citep{joh08} that suggests stream-like substructures are rare
in inner halo.  If ECHOS really are shell-like substructures on radial
orbits as opposed to stream-like substructures on tangential orbits as
proposed by \citetalias{schl09}, measurement over a small patch of sky
might still intersect orbits in a range of orbital phase and therefore
produce large velocity dispersions that are not representative of the
velocity dispersion of the progenitor.  We argue that this is unlikely,
however, as that would imply that the radial velocity dispersion
within a single ECHOS is a function of apparent magnitude, a trend not
observed in ECHOS (see Figures 2 through 11 in \citetalias{schl09}).
With those caveats in mind, typical classical dwarf spheroidal (dSph)
galaxies have radial velocity dispersions $\sigma \sim 10$ km s$^{-1}$
\citep[e.g.,][]{mat98}.  Indeed, three of our ECHOS even have velocity
dispersions comparable to that observed in the Small Magellanic Cloud
(SMC) in which \citet{har06} measured $\sigma \approx 27.5 \pm 0.5$
km s$^{-1}$.  The median radial velocity dispersion of ECHOS $\sigma
\sim 20$ km s$^{-1}$ is also characteristic of a dSph like NGC 147
or NGC 185.  This observation implies an even more massive progenitor
though, as the fact that ECHOS are not associated with surface
brightness substructure suggests that the velocity dispersion has had
time to cool measurably.

The observation that ECHOS are less $\alpha$-enhanced than typical
inner halo stars suggests that the star formation timescale in
the ECHOS progenitor was long relative to the star formation
timescale in the massive progenitor of the bulk of the inner halo
\citep[e.g.,][]{mcw95}, so long as the stellar initial mass function is
not too different in the two environments.  This relatively extended
star formation timescale might disfavor globular clusters as ECHOS
progenitors, though some globular clusters have [$\alpha$/Fe] $\sim
0.2$ \citep[e.g.,][]{carn96,brod06,carr09}.  On the other hand, the
characteristic ECHOS [Fe/H] $\sim -1.0$ falls between the two peaks in
the observed bimodal Milky Way globular cluster [Fe/H] distribution
\citep[e.g.,][]{arm88,brod06}.  Additionally, globular clusters that
metal-rich are very rare in the inner halo \citep[e.g.,][]{gei07}.
In any case, the large radial velocity dispersion of ECHOS indicates
that ECHOS are unlikely to be the debris of tidally disrupted globular
clusters, as globular clusters have velocity dispersions $\sigma \sim 5$
km s$^{-1}$ \citep[e.g.,][]{man91}.  Core-collapse globular clusters
have higher velocity dispersions, but are also resistant to disruption.

One last possibility is that the stars in ECHOS formed in the nascent
disk of the Milky Way and were scattered into the inner halo during an
accretion event.  Recent cosmological models of Milky Way formation
by \citet{zol09} suggest that scattering into the inner halo does
contribute to the stellar population in the inner halo.  Moreover,
the accretion of dSph galaxies more massive then those suggested by
the characteristic ECHOS [Fe/H] $\sim -1.0$ can scatter substantial
numbers of disk stars into the inner halo \citep[e.g.,][]{pur10}.
However, models of the chemical evolution of the disk of the Milky Way
suggest that the typical chemical composition of ECHOS is difficult
to explain with a disk population \citep{sch09a,sch09b}.  In addition,
\citet{zol10} recently suggested that disk stars scattered into the halo
should be more $\alpha$-enhanced than inner halo stars at constant [Fe/H].
Improved resolution and better treatments of star formation and feedback
in theoretical models may yet determine whether scattering of disk stars
into the inner halo by low-mass accretion events is a plausible origin
of ECHOS.

\section{Conclusion}

The metal-poor main sequence turnoff stars in the elements of cold
halo substructure in the inner halo of the Milky Way found to be
kinematically distinct from the kinematically smooth inner halo population
by \citet{schl09} are also chemically distinct from the smooth inner halo.
As a population, ECHOS are more iron-rich and less $\alpha$-enhanced than
the MPMSTO stars in the kinematically smooth halo population along the
same line of sight.  ECHOS are chemically distinct from both the thin and
thick disk as well, as they are more iron-poor than average thick-disk
stars and both more iron-poor and more $\alpha$-enhanced than average
thin-disk stars.  Kinematically, the typical velocity dispersion of the
ECHOS population is $\sigma \sim 20$ km s$^{-1}$, though ECHOS have a
range of radial velocity dispersions extending from the floor of $\sigma
\sim 10$ km s$^{-1}$ set by the precision of SEGUE radial velocities
at the apparent magnitude of the stars in ECHOS to the maximum observed
value of $\sigma \sim 30$ km s$^{-1}$.  If ECHOS are the result of the
tidal disruption of an accreted dwarf spheroidal galaxy or galaxies,
the high iron metallicity [Fe/H] $\sim -1.0$, large velocity dispersion
$\sigma \sim 20$ km s$^{-1}$, and the lack of corresponding surface
brightness substructure imply the accretion of a $M_{\mathrm{tot}}
\gtrsim 10^9~M_{\odot}$ halo sometime in the last 5 Gyr since $z=0.5$.
In addition, the high iron metallicities, low $\alpha$-enhancements,
and large velocity dispersions are difficult to reconcile with globular
clusters as the progenitors of ECHOS.  One final possible explanation
for these observations is that the stars in ECHOS formed in the nascent
disk of the Milky Way and were subsequently scattered into the inner
halo by cosmologically common low-mass accretion events, though models
of the chemical evolution of the Milky Way's disk cast some doubt on
this scenario as well.

\acknowledgments We thank James Bullock, Puragra Guhathakurta, Evan
Kirby, Piero Madau, Sarah Martell, Ralph Sch{\"o}nrich, and Matthias
Steinmetz for useful comments and conversation.   We also thank the
anonymous referee for suggestions that improved the clarity of this
manuscript.  This research has made use of NASA's Astrophysics Data System
Bibliographic Services.  This material is based upon work supported under
a National Science Foundation Graduate Research Fellowship.  Y.S.L. and
T.C.B. acknowledge partial support for this work from PHY 02-16783 and
PHY 08-22648: Physics Frontiers Center / Joint Institute for Nuclear
Astrophysics (JINA), awarded by the U.S. National Science Foundation.
Funding for the SDSS and SDSS-II has been provided by the Alfred
P. Sloan Foundation, the Participating Institutions, the National Science
Foundation, the U.S. Department of Energy, the National Aeronautics and
Space Administration, the Japanese Monbukagakusho, the Max Planck Society,
and the Higher Education Funding Council for England. The SDSS Web Site
is http://www.sdss.org/.

The SDSS is managed by the Astrophysical Research Consortium for the
Participating Institutions. The Participating Institutions are the
American Museum of Natural History, Astrophysical Institute Potsdam,
University of Basel, University of Cambridge, Case Western Reserve
University, University of Chicago, Drexel University, Fermilab, the
Institute for Advanced Study, the Japan Participation Group, Johns
Hopkins University, the Joint Institute for Nuclear Astrophysics, the
Kavli Institute for Particle Astrophysics and Cosmology, the Korean
Scientist Group, the Chinese Academy of Sciences (LAMOST), Los Alamos
National Laboratory, the Max-Planck-Institute for Astronomy (MPIA),
the Max-Planck-Institute for Astrophysics (MPA), New Mexico State
University, Ohio State University, University of Pittsburgh, University
of Portsmouth, Princeton University, the United States Naval Observatory,
and the University of Washington.

{\it Facilities:} \facility{Sloan}

\appendix

\section{A. Detailed Description of Coaddition Algorithm} 

The selection of the input sample of spectra to the coaddition process
is different in each of the three contexts described in Sections 3.3,
3.4, and 3.5.  In all cases, we select a bootstrap sample of spectra,
coadd the spectra in the bootstrap sample into a single coadded spectrum,
and then use the SSPP to analyze the coadded spectrum.  That analysis
produces a single estimate of the mean metallicity of the population of
stars from which the bootstrap sample was drawn.  We draw many bootstrap
samples, derive many estimates of the mean metallicity of the population,
and use that distribution to quantify both the mean metallicity of
the population and our uncertainty in that estimate.

In Section 3.3, we estimate the accuracy and precision of the SSPP on
coadded spectra in a range of metallicity ($-2.41 < \mathrm{[Fe/H]} <
-0.23$) and S/N ($10 \lesssim \mathrm{S/N} \lesssim 100$).  For each
of eight MPMSTO stars with metallicity known from high resolution
spectroscopy, the input sample is a collection of 640 spectra produced
by adding noise to the original high S/N SEGUE spectrum of that star.
There are no radial velocity or effective temperature cuts applied in
this case.

In Section 3.4, we estimate the accuracy and precision of the SSPP
on coadded spectra produced by bootstrap sampling from the MPMSTO
populations of the globular clusters M~13 and M~15.  In this case, though
the individual spectra belong to stars with very similar metallicities,
the spectra will correspond to stars with different radial velocities and
effective temperature.  As a result, we can use the spectra associated
with M~13 and M~15 MPMSTO stars to quantify the effect of coadding stars
in a finite range of effective temperature, surface gravity, and radial
velocity.  Specifically, we include in the sample of input spectra only
those spectra corresponding to MPMSTO stars with equatorial coordinates
within the projected tidal radius of each cluster, with radial velocities
close to the systematic radial velocity of the cluster, and within the
500 K range of effective temperature that produces the highest S/N in
the final coadded spectrum.

In Section 3.5, we use our algorithm to estimate both the average
metallicity of MPMSTO stars in ECHOS and in the kinematically smooth
component of the halo along the same line of sight.  We include in
each ECHOS sample of input spectra only those spectra corresponding
to MPMSTO stars with radial velocities within the radial velocity
overdensity that defines the ECHOS (as defined in \citetalias{schl09}
and reported in Tables~\ref{tbl-1}, \ref{tbl-2}, and \ref{tbl-3}).
We include all other MPMSTO spectra obtained along the line of sight in
the kinematically smooth halo input sample.  In both cases, we use the
same effective temperature criteria used in the globular cluster case.

The following eight steps describe how a single bootstrap coadded spectrum
can be produced from the input samples described above.

\begin{enumerate}
\item
Use resampling with replacement to select $N$ spectra from the $N$
available spectra to create a bootstrap sample of spectra to synthesize
into a single coadd spectrum.
\item
Let $f_{i,j}(\lambda_{j}')$ and $g_{i,j}(\lambda_{j}')$ denote the flux
and inverse variance from the SDSS spectroscopic reduction pipeline.
The subscript $i$ refers to the $i$th star in the coadd and the subscript
$j$ refers to the $j$th wavelength bin.
\item
Shift each spectrum to redshift $z = 0$ such that

\begin{eqnarray} \label{Aeq1}
\lambda_{j} & = & \frac{\lambda_{j}'}{1+z}
\end{eqnarray}

\noindent
and denote the radial-velocity zeroed spectrum and inverse variance
$f_{i,j}(\lambda_{j})$ and $g_{i,j}(\lambda_{j})$ respectively.
\item
Use natural cubic spline interpolation to interpolate
$f_{i,j}(\lambda_{j}) \rightarrow f_{i}(\lambda)$ and
$g_{i,j}(\lambda_{j}) \rightarrow g_{i}(\lambda)$.
\item
Sample the continuous functions $f_{i}(\lambda)$ and $g_{i}(\lambda)$
on to a common grid in $\lambda_{k}$ for every star from $\lambda_{1}
= 3850$ \AA~to $\lambda_{2} = 9000$ \AA~in $n$ 0.5~\AA~increments.
Set any $f_{i,k}(\lambda_{k}) < 0$ to $f_{i,k}(\lambda_{k}) = 0$ and
any $g_{i,k}(\lambda_{k}) \leq 0$ to $g_{i,k}(\lambda_{k}) = 10^{-6}$.
\item
To this point, each star has a different total flux in its spectrum,
so rescale $f_{i}$ and $g_{i}$ to the same total flux while conserving
the S/N ratio in each bin.  In other words,

\begin{eqnarray} \label{Aeq2}
A_{i} & = & \int_{\lambda_{1}}^{\lambda_{2}} f_{i}(\lambda) d\lambda \\
f_{i,k}(\lambda_{k}) & = & \frac{1}{A_i} f_{i,k}(\lambda_{k}) \\
g_{i,k}(\lambda_{k}) & = & A_{i}^{2} g_{i,k}(\lambda_{k}).
\end{eqnarray}
\item
Loop over all $n$ wavelength bins to create the coadd spectrum and its inverse
variance using the weighted-mean such that

\begin{eqnarray} \label{Aeq3}
w_{i} & = & \left\{g_{1,k}(\lambda_{k}),g_{2,k}(\lambda_{k}),\ldots,
g_{N,k}(\lambda_{k})\right\} \\
\overline{f}_{k}(\lambda_{k}) & = &
\frac{\sum_{i=1}^{N} w_{i} f_{i,k}(\lambda_{k})}{\sum_{i=1}^{N} w_{i}} \\
\overline{g}_{k}(\lambda_{k}) & = & \sum_{i=1}^{N} w_{i}.
\end{eqnarray}
\item
The coadded spectrum $\overline{f}_{k}(\lambda_{k})$ and its inverse
variance $\overline{g}_{k}(\lambda_{k})$ are then processed by the SSPP
to produce a single estimate of the mean metallicity of the input sample.
\end{enumerate}

\begin{figure}
\plottwo{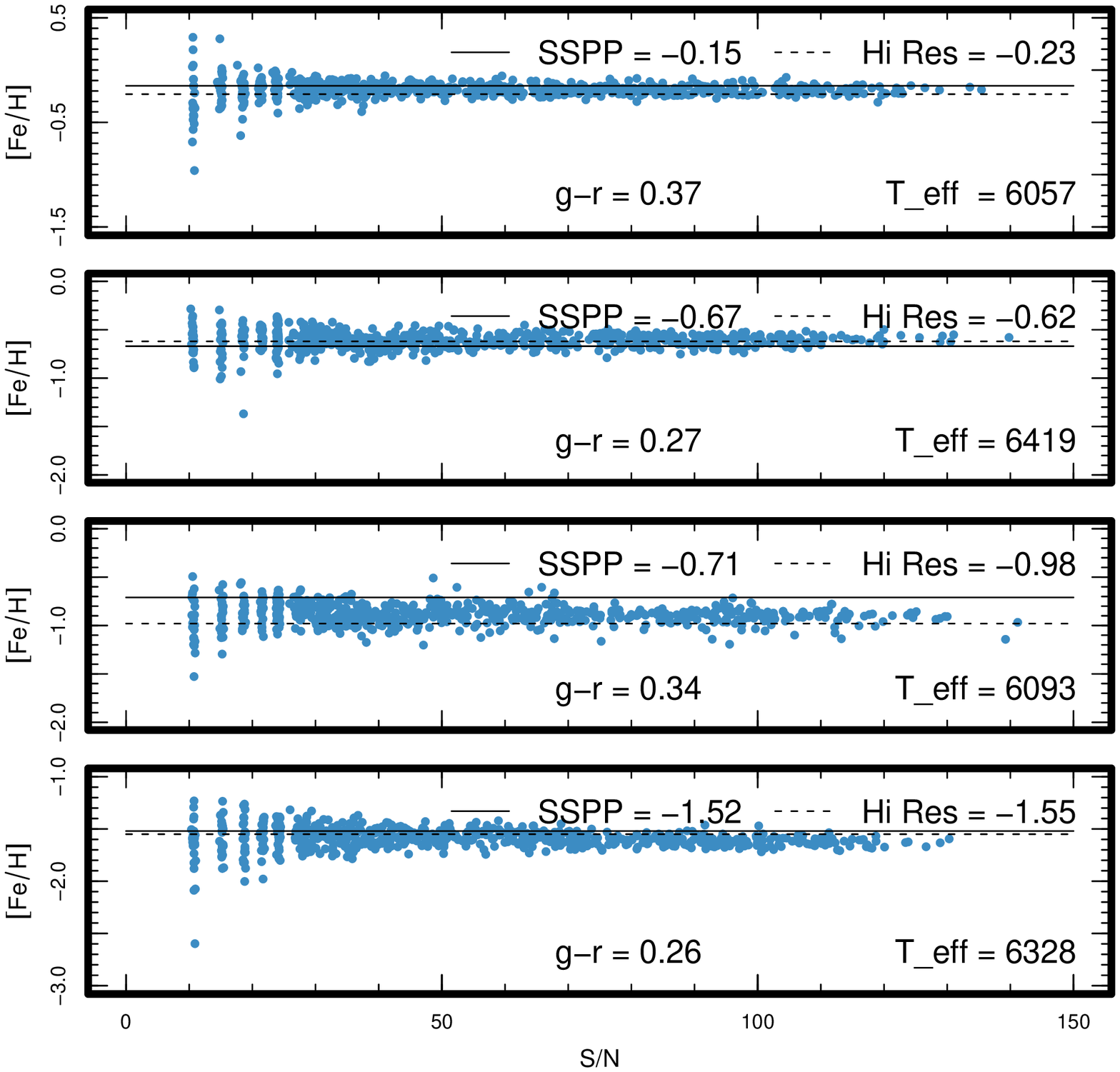}{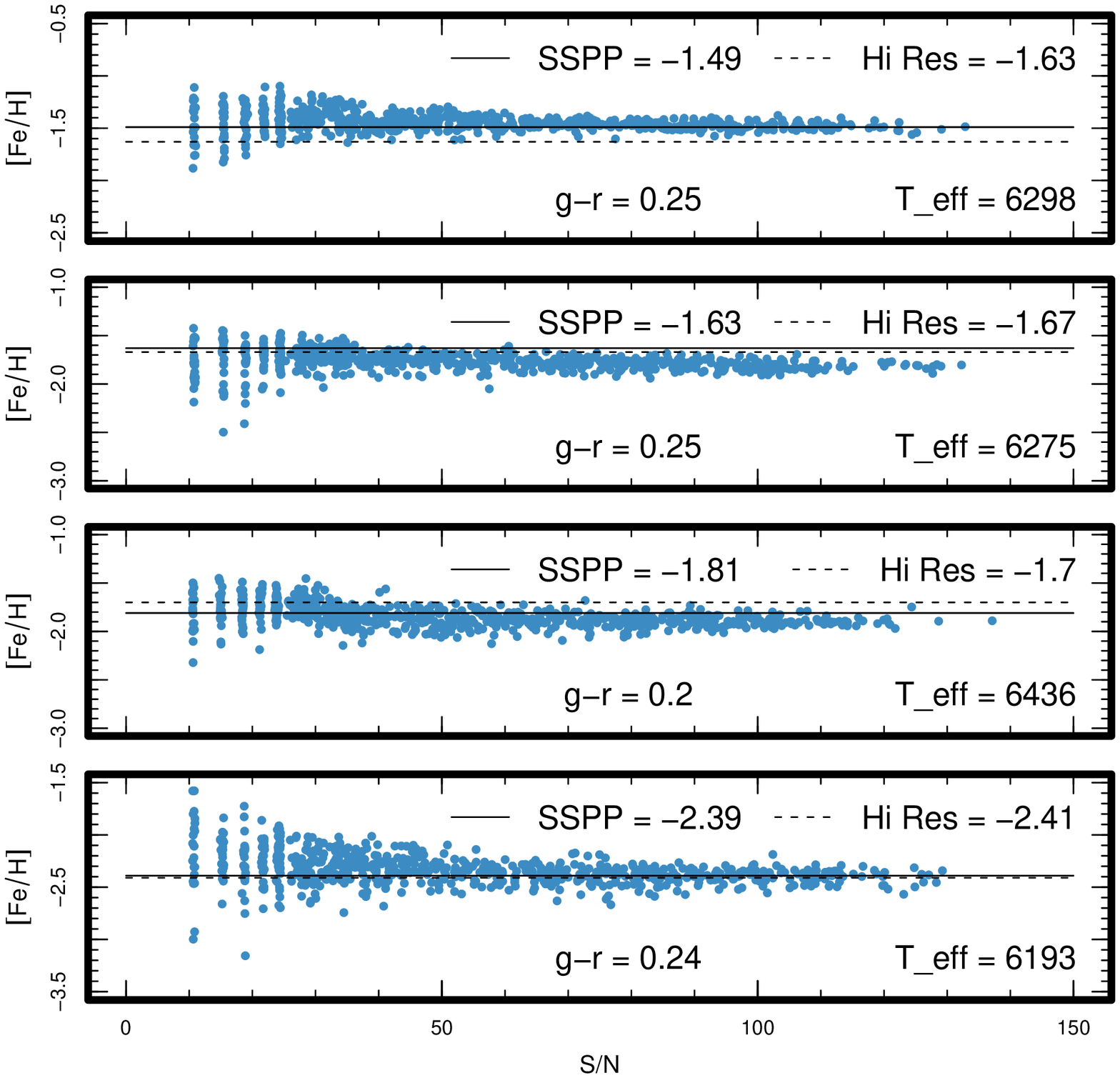}
\caption{Ensemble of bootstrap coadd realizations of single MPMSTO
stars giving SSPP [Fe/H] precision and accuracy as a function of S/N
for that star.  The [Fe/H] and $T_{\mathrm{eff}}$ estimates given for
each star were derived from Hobby-Eberly Telescope High Resolution
Spectrograph observations \citep{all08}; we also include the SSPP
[Fe/H] estimate for the original, single high S/N spectrum.  Each point
in the plot corresponds to the result of a SSPP analysis of a single
bootstrap coadded spectrum made up of many model low S/N MPMSTO star
spectra created by degrading a high S/N SEGUE MPMSTO star spectrum with
a detailed noise model.  The mean square error (MSE $\equiv$ bias$^2$ +
variance) of our SSPP analysis of codded spectra ranges from 0.05 dex in
[Fe/H] for the most iron-rich stars to 0.20 dex in [Fe/H] for the most
iron-poor stars.\label{fig01}}
\end{figure}

\begin{figure}
\plottwo{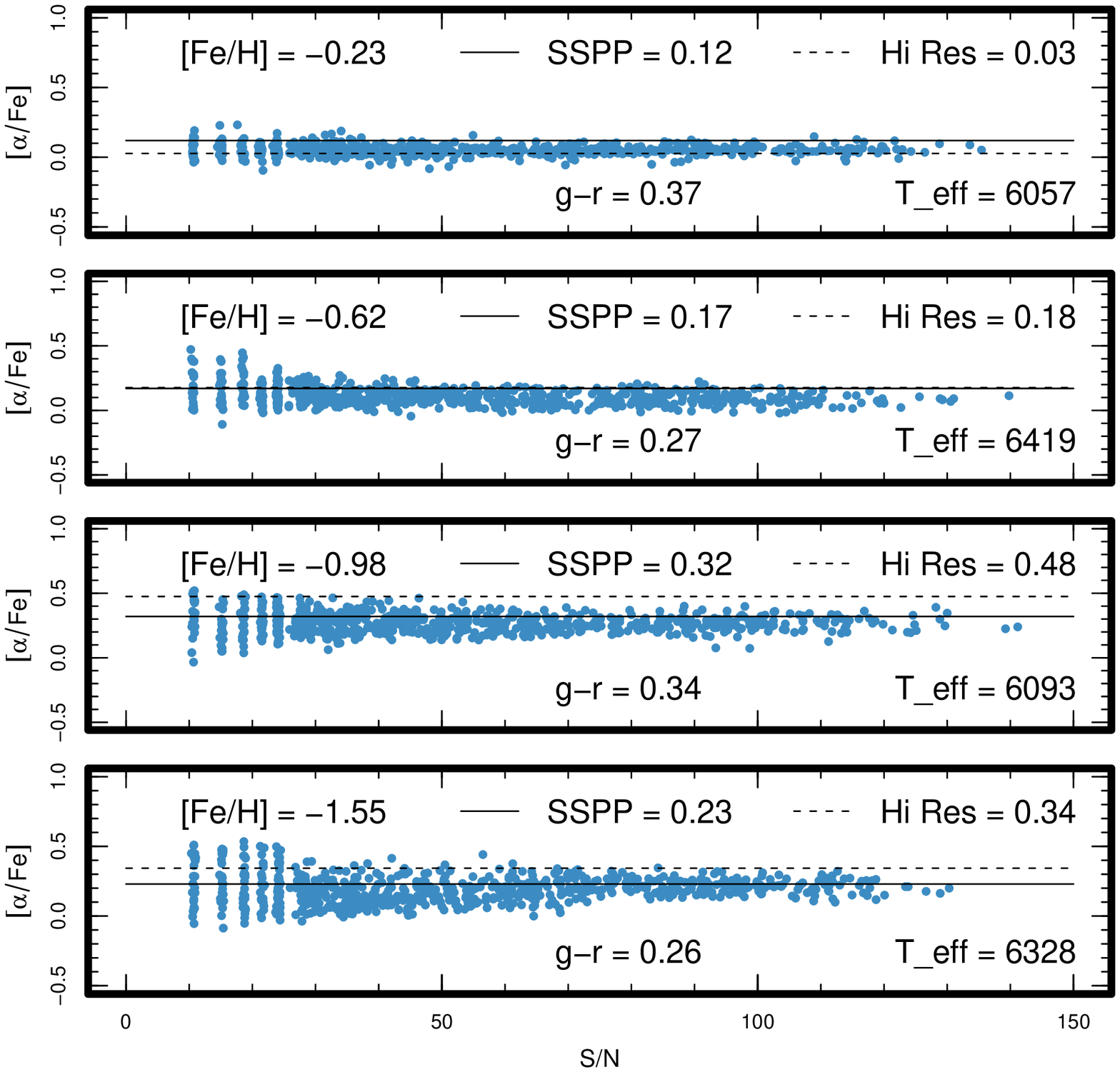}{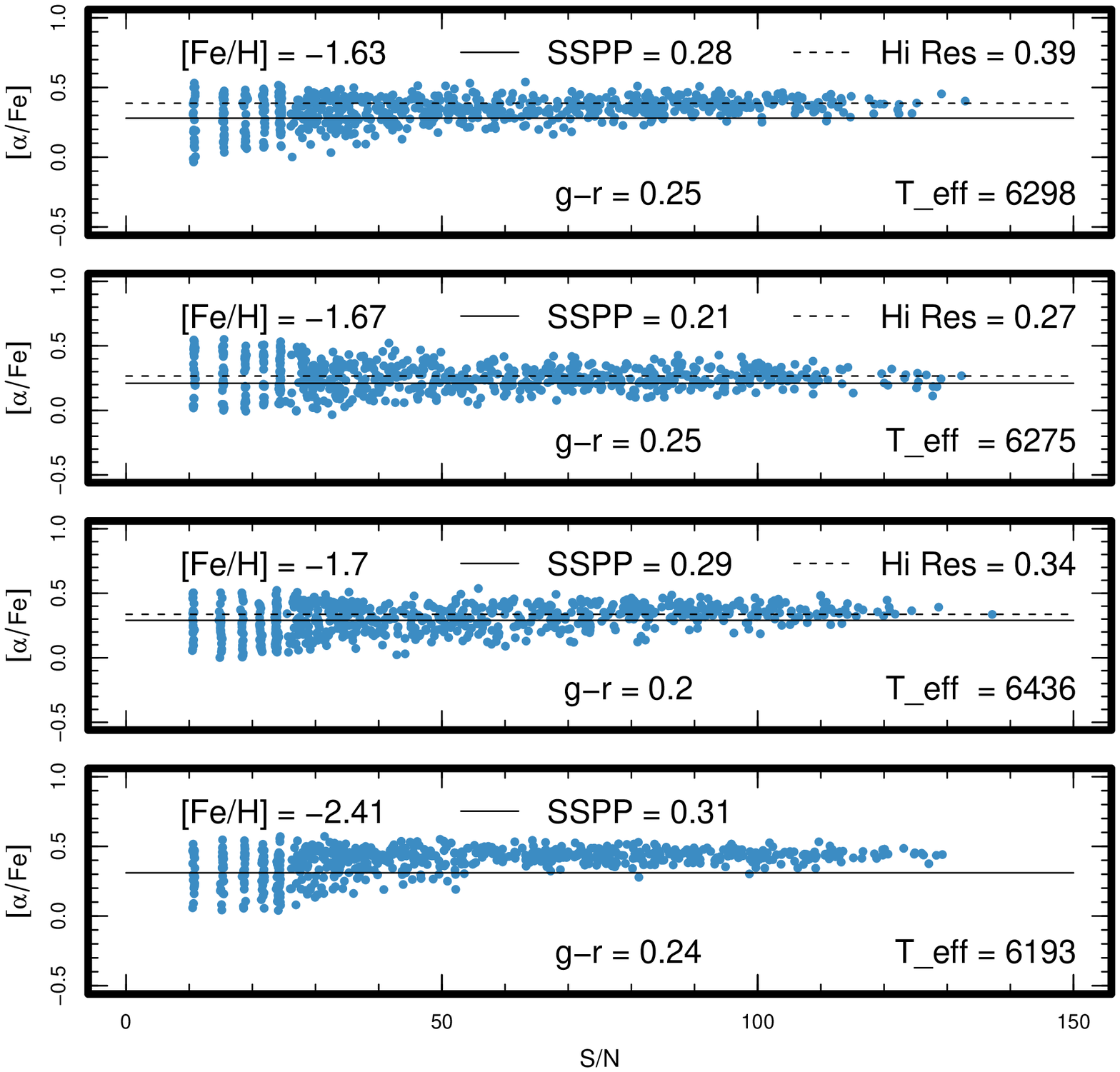}
\caption{Ensemble of bootstrap coadd realizations of single MPMSTO stars
giving SSPP [$\alpha$/Fe] precision and accuracy as a function of S/N
for that star.  The panels are in the same order as in Figure~\ref{fig01}.
The [$\alpha$/Fe] and $T_{\mathrm{eff}}$ estimates given for each star
were derived from Hobby-Eberly Telescope High Resolution Spectrograph
observations \citep{all08}; we also include the SSPP [$\alpha$/Fe]
estimate for the original, single high S/N spectrum.  Each point in
the plot corresponds to the result of a SSPP analysis of a single
bootstrap coadded spectrum made up of many model low S/N MPMSTO star
spectra created by degrading a high S/N SEGUE MPMSTO star spectrum with
a detailed noise model.  The MSE of our SSPP analysis of codded spectra
ranges from 0.05 dex in [$\alpha$/Fe] for the most iron-rich stars to
0.20 dex in [$\alpha$/Fe] for the most iron-poor stars.\label{fig02}}
\end{figure}

\begin{figure}
\epsscale{0.8}
\plotone{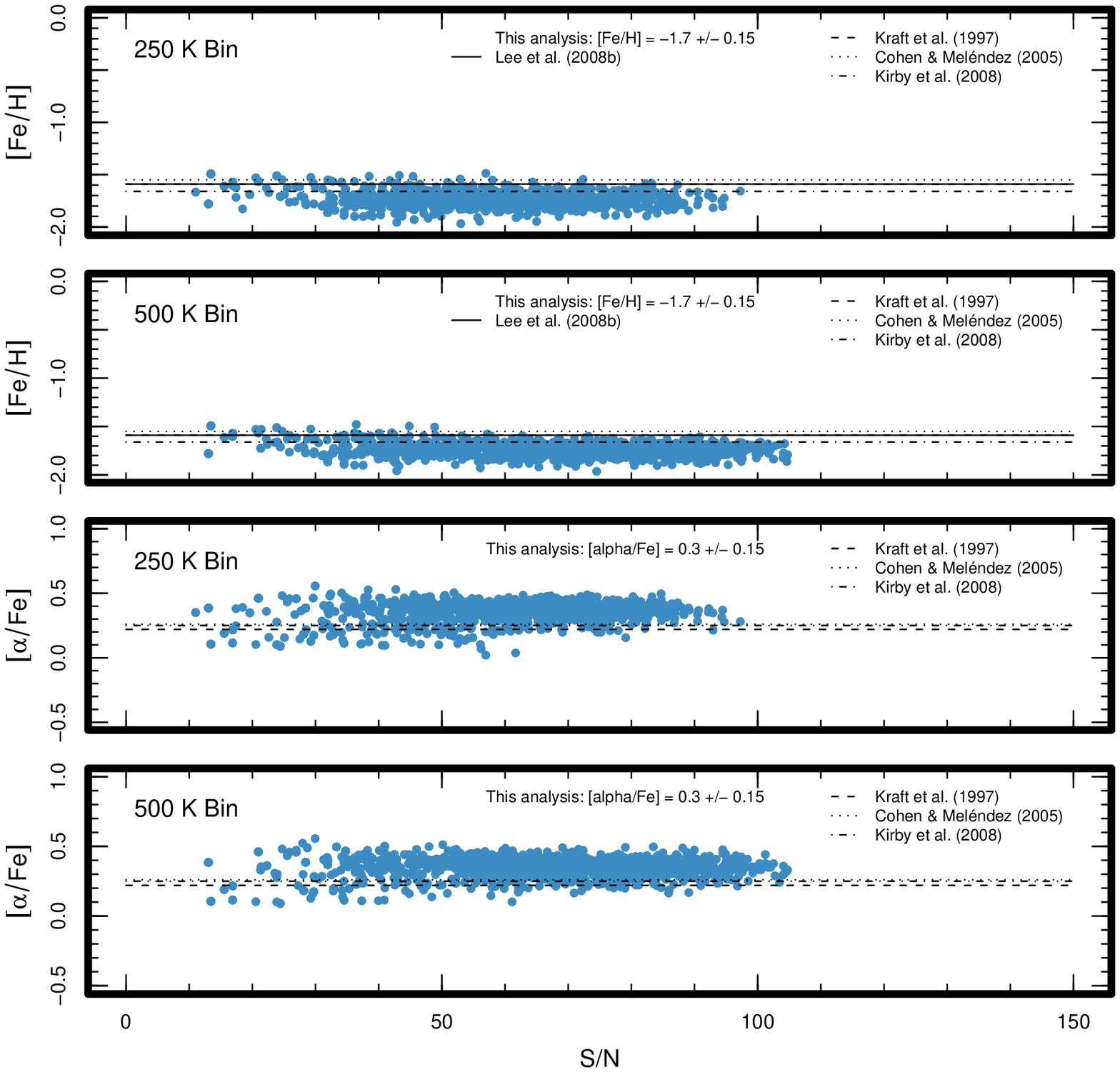}
\caption{Ensemble of bootstrap coadd realizations of single MPMSTO
stars giving SSPP [Fe/H] and [$\alpha$/Fe] precision and accuracy as
a function of S/N for M~13.  Each point in the plot corresponds to a
single bootstrap coaddition of SEGUE MPMSTO spectra corresponding to
stars with equatorial coordinates within the projected tidal radius of
M~13, with radial velocities within 15 km s$^{-1}$ of the systematic
radial velocity of the cluster, and with $g-r$ colors in an interval that
corresponds to a range in effective temperature as noted in each panel.
Note that the precision and accuracy of the SSPP analysis of the coadded
spectra is not a strong function of the $g-r$ color range (and therefore
$T_{\mathrm{eff}}$ range) of the MPMSTO stars included in the coadd.
Combined with the fact that coadding in 500 K bins always leads to a
higher S/N coadded spectrum, we coadd spectra corresponding to MPMSTO
stars in the bin of $g-r$ color corresponding to the $T_{\mathrm{eff}}$
bin of width 500 K centered on the $T_{\mathrm{eff}}$ value that gives
the highest S/N in the resultant coadded spectrum.  In this way, the
SSPP produces an estimate for the metallicity of M~13 of [Fe/H] $=
-1.7 \pm 0.15$ and [$\alpha$/Fe] $= 0.3 \pm 0.15$ in rough agreement
with previous analyses \citep{kra97,coh05,kir08a}.\label{fig03}}
\end{figure}

\begin{figure}
\epsscale{1.0}
\plotone{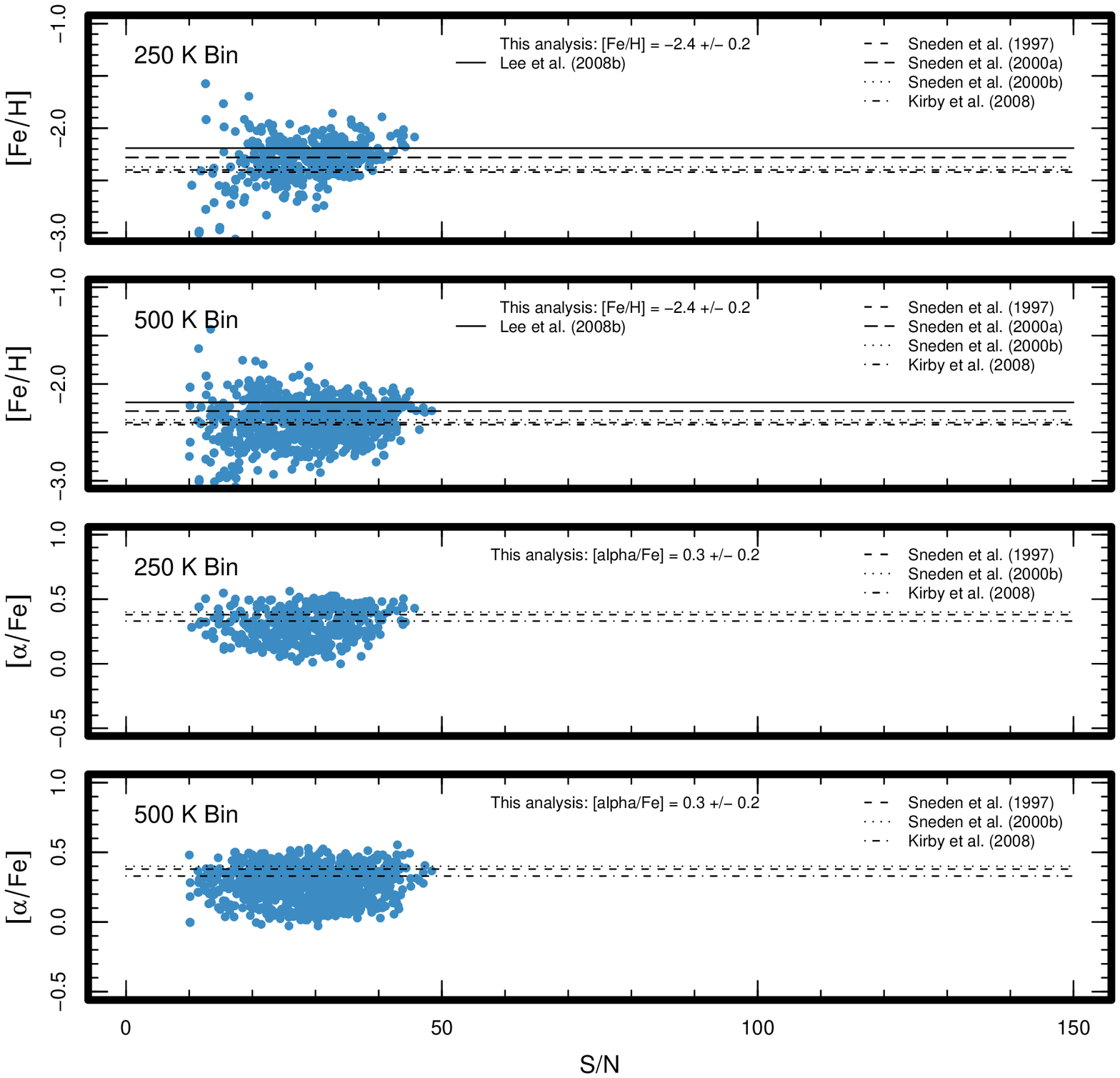}
\caption{Ensemble of bootstrap coadd realizations of single MPMSTO
stars giving SSPP [Fe/H] and [$\alpha$/Fe] precision and accuracy as a
function of S/N for M~15.  Each point in the plot corresponds to a single
bootstrap coaddition of SEGUE MPMSTO spectra corresponding to stars with
equatorial coordinates within the projected tidal radius of M~15, with
radial velocities within 25 km s$^{-1}$ of the systematic radial velocity
of the cluster, and with $g-r$ colors in an interval that corresponds to
a range in effective temperature as noted in each panel.  In this way,
the SSPP produces an estimate for the metallicity of M~15 of [Fe/H] $=
-2.4 \pm 0.2$ and [$\alpha$/Fe] $= 0.3 \pm 0.2$ in rough agreement with
previous analyses \citep{sne97,sne00a,sne00b,kir08a}.\label{fig04}}
\end{figure}

\begin{figure}
\plottwo{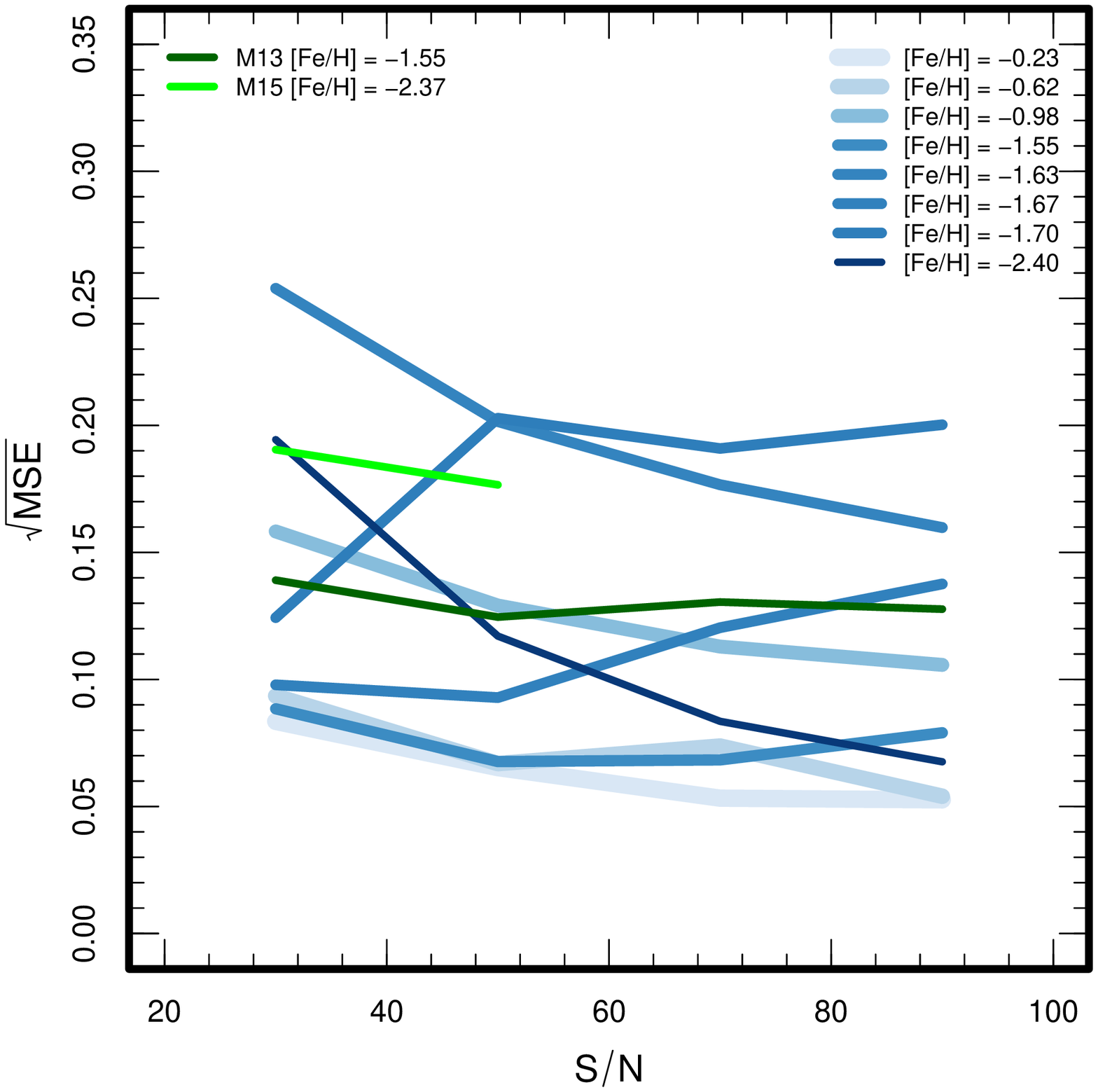}{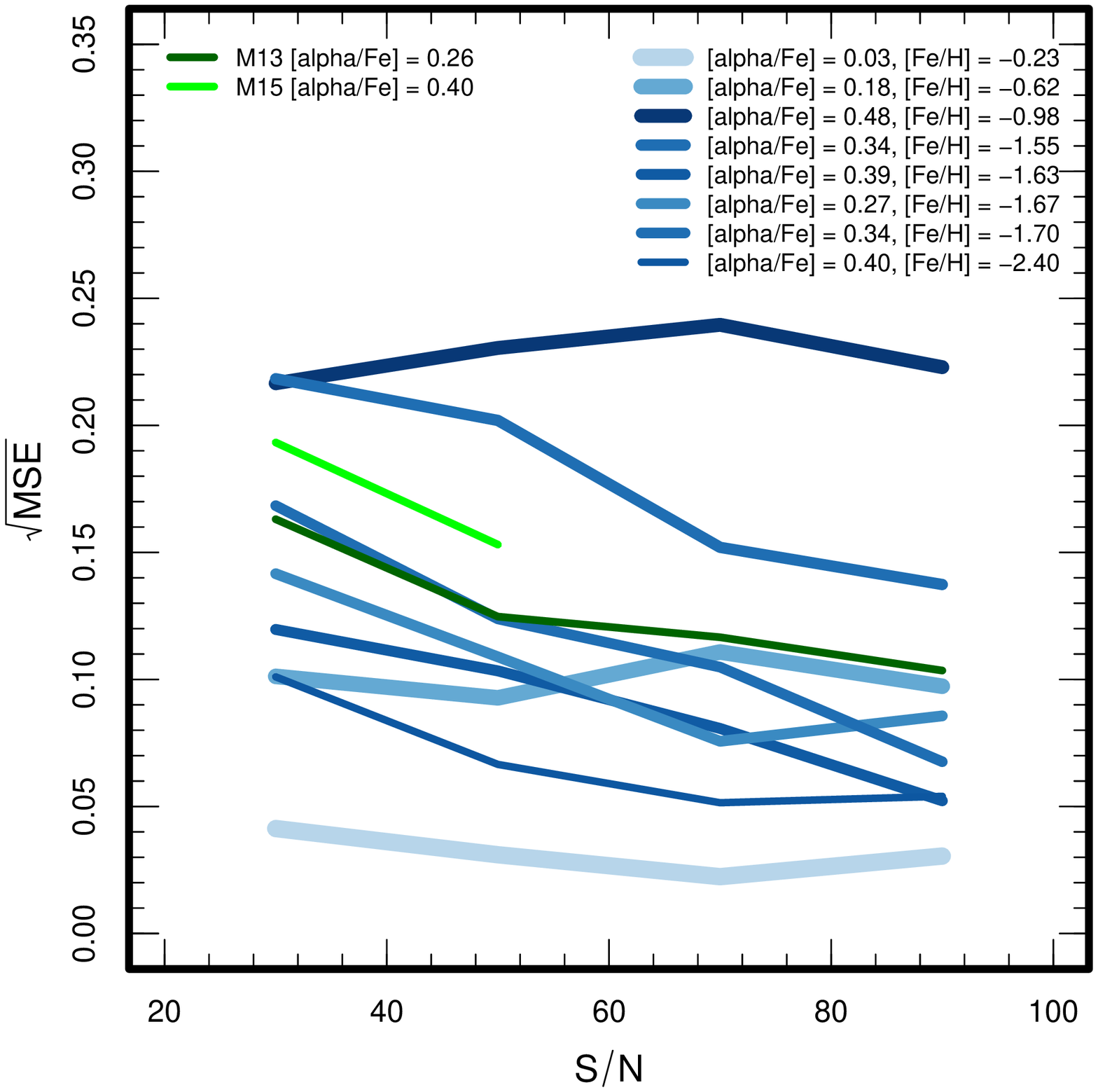}
\caption{The mean square error (MSE $\equiv$ bias$^2$ + variance)
of our SSPP analysis of coadded MPMSTO spectra as a function of S/N
and metallicity.  The blue curves correspond to our analysis of the
noise-added spectra, while the green curves correspond to our analysis
for M~13 and M~15.  The widths of the curves are common between both
panels in that the curve for a single star is plotted with the same width
in both panels.  In both panels, the most solar-like compositions are
always the lightest shades of blue.  Note that the SSPP produces more
precise [Fe/H] and [$\alpha$/Fe] estimates at common S/N for the most
metal-rich MPMSTO stars.  In that case, the abundance of metal lines
in the moderate resolution SEGUE spectra of metal-rich stars permits
precise [Fe/H] estimates.  Also note that the MSE of our analysis is
more strongly affected by population metallicity than it is affected by
the coaddition of MPMSTO star spectra with a small but finite range of
$T_{\mathrm{eff}}$ and $\log{g}$.  For that reason, the expected MSE
of our analysis is best determined by our analysis of the noise-added
spectra.  Ultimately, including both statistical and systematic effects,
our SSPP analysis of coadded SEGUE MPMSTO spectra produces estimates
that are precise and accurate enough to identify differences on the
order of 0.2 dex in both [Fe/H] and [$\alpha$/Fe].  \emph{Left}: SSPP
[Fe/H] MSE as a function of S/N and metallicity.  \emph{Right}: SSPP
[$\alpha$/Fe] MSE as a function of S/N and metallicity.\label{fig05}}
\end{figure}

\begin{figure}
\epsscale{0.8}
\plottwo{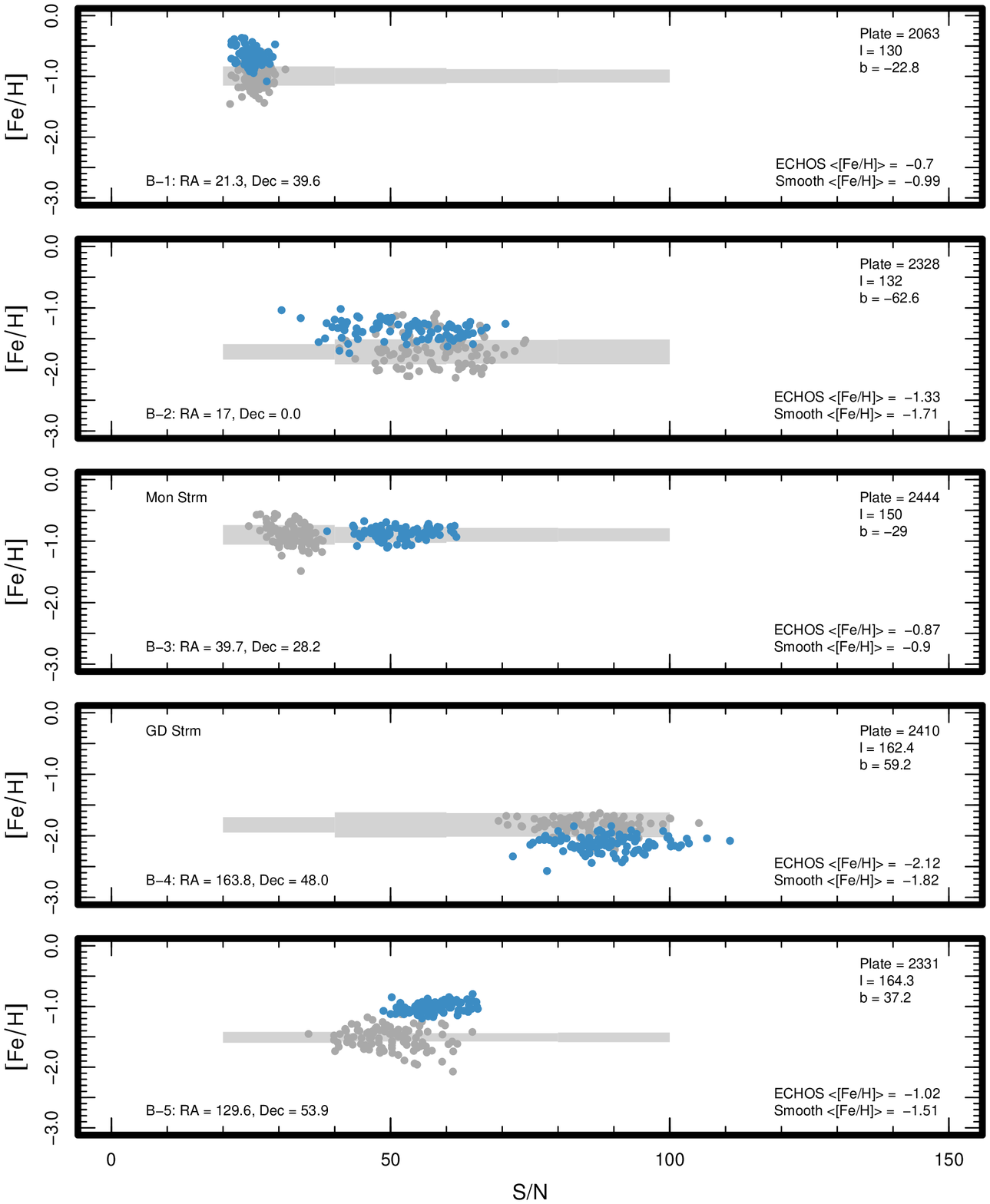}{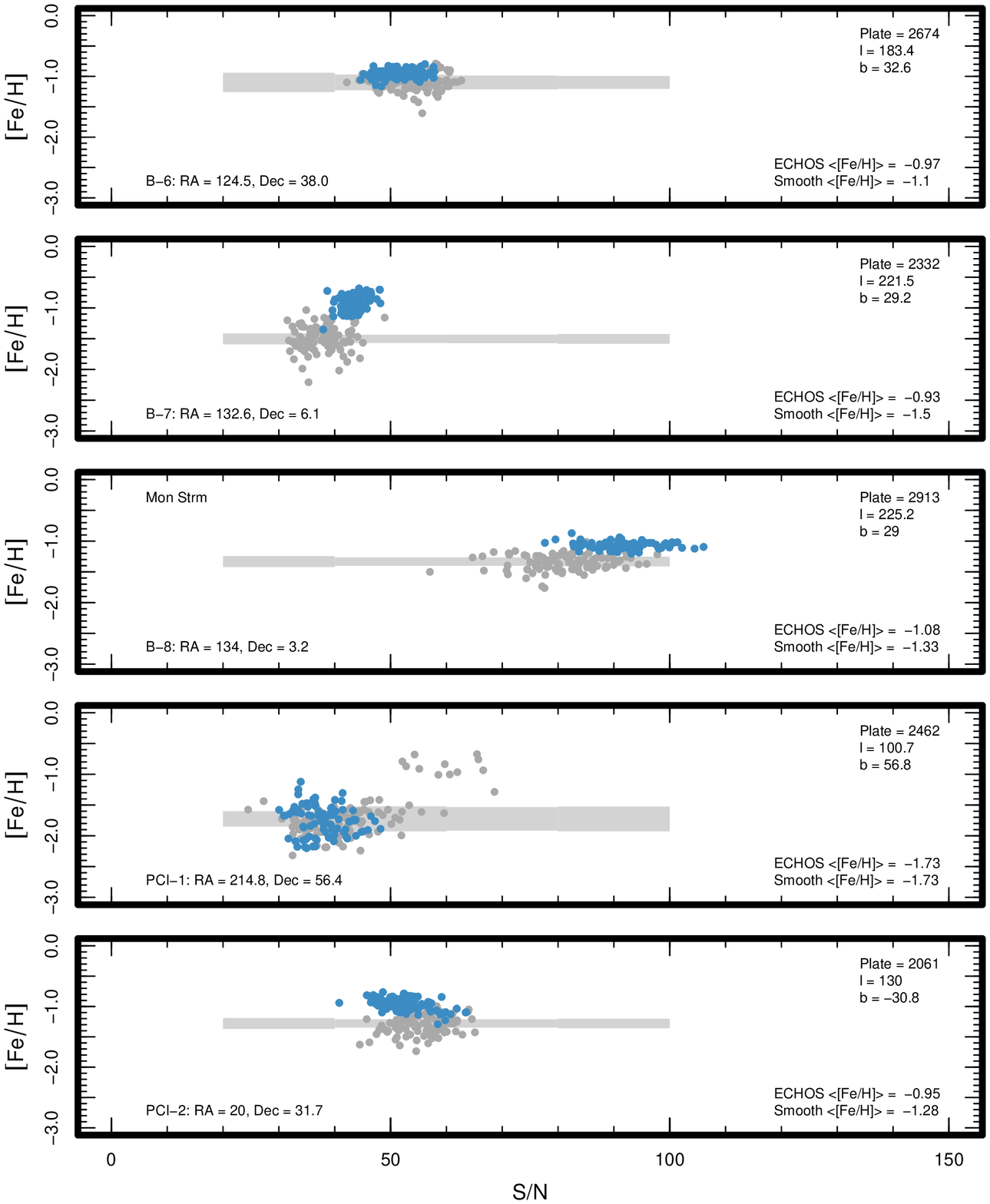}
\caption{Ensemble of bootstrap coadd realizations giving SSPP [Fe/H]
estimates for ECHOS from \citetalias{schl09} using bootstrap coaddition
of individual SEGUE MPMSTO star spectra.  Each blue point corresponds
to a SSPP [Fe/H] estimate for a single ECHOS bootstrap coadd, while the
dark gray points are SSPP [Fe/H] estimates from an equivalent bootstrap
coaddition of MPMSTO star spectra with radial velocities indicating
membership in the kinematically smooth inner halo MPMSTO population
along the same SEGUE line of sight that hosts the ECHOS.  MPMSTO stars
correspond to an ECHOS if they have a radial velocity consistent with
radial velocity overdensities as determined by \citetalias{schl09};
we identify all other stars along the same line of sight as the
kinematically smooth inner halo MPMSTO population.  Of those MPMSTO
stars that are radial velocity members of an ECHOS, we include in the
coadd only those spectra that correspond to stars that have $g-r$
colors corresponding to the $T_{\mathrm{eff}}$ bin of width 500 K
centered on the $T_{\mathrm{eff}}$ value that gives the highest S/N in
the resultant coadded spectrum.  In the same way, we choose an optimal
$T_{\mathrm{eff}}$ bin of width 500 K from which we select spectra for
coaddition from the kinematically smooth inner halo MPMSTO population.
The light gray polygon is centered on the mean metallicity of the smooth
component of the inner halo along each SEGUE line of sight and indicates
the expected MSE of the SSPP at the metallicity of the smooth component
as a function of S/N from our bootstrap coaddition of high S/N spectra
degraded with the noise model (see Figure~\ref{fig05}).  Therefore,
if the cloud of blue points does not coincide with the gray polygon,
then the ECHOS is chemically distinct from the kinematically smooth
population along that line of sight.  In all cases where the quoted
metallicity of the smooth component is more iron-rich than typically
associated with the smooth inner halo, the reason is because the ECHOS
dominates the MPMSTO population along that line of sight (see Figures 2
through 11 of \citetalias{schl09}).  \emph{Left}: SSPP [Fe/H] analysis
for ECHOS B-1, B-2, B-3 (Monoceros), B-4 \citep{gri06b}, and B-5 from
Table~\ref{tbl-1}.  \emph{Right}: SSPP [Fe/H] analysis for ECHOS B-6,
B-7, and B-8 from Table~\ref{tbl-1} as well as PCI-1 and PCI-2 from
Table~\ref{tbl-2}.\label{fig06}}
\end{figure}

\begin{figure}
\epsscale{1.0}
\plottwo{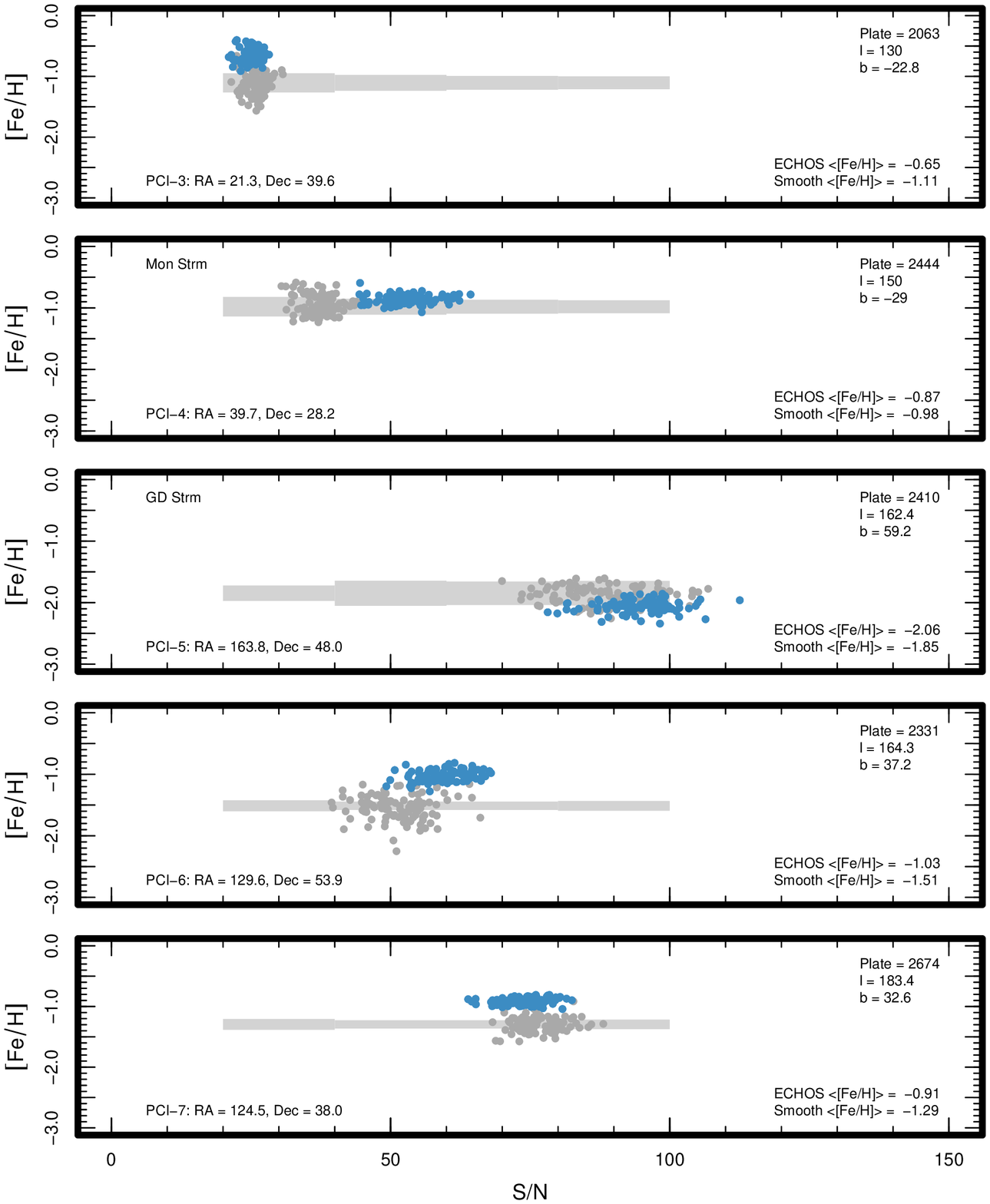}{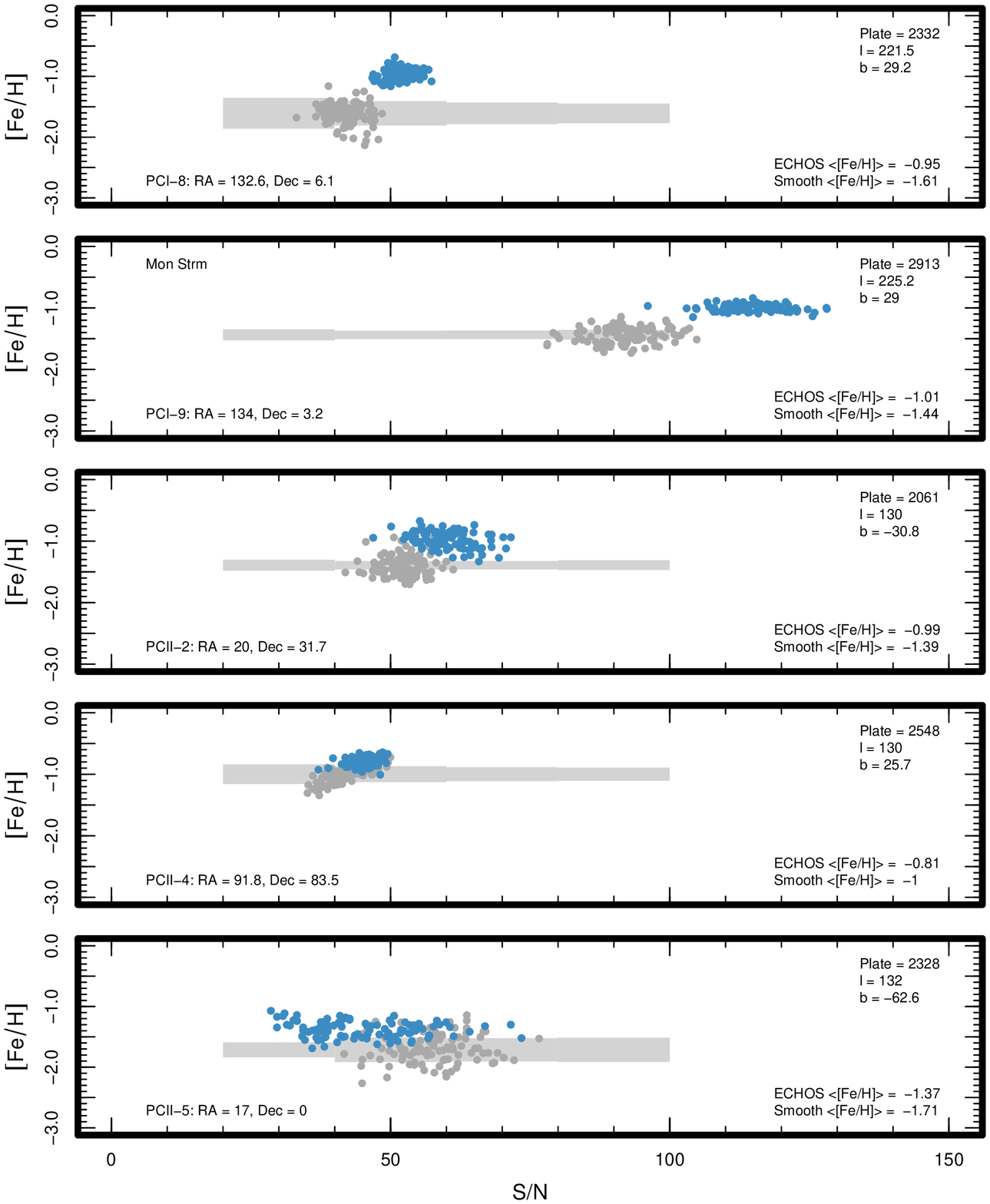}
\caption{\emph{Left}: SSPP [Fe/H] analysis for ECHOS PCI-3,
PCI-4 (Monoceros), PCI-5 \citep{gri06b}, PCI-6, and PCI-7 from
Table~\ref{tbl-2}.  \emph{Right}: SSPP [Fe/H] analysis for ECHOS PCI-8
and PCI-9 (Monoceros) from Table~\ref{tbl-2} as well as PCII-2, PCII-4,
and PCII-5 from Table~\ref{tbl-3}.  See the caption to Figure~\ref{fig06}
for a detailed description of this type of figure.\label{fig07}}
\end{figure}

\begin{figure}
\plottwo{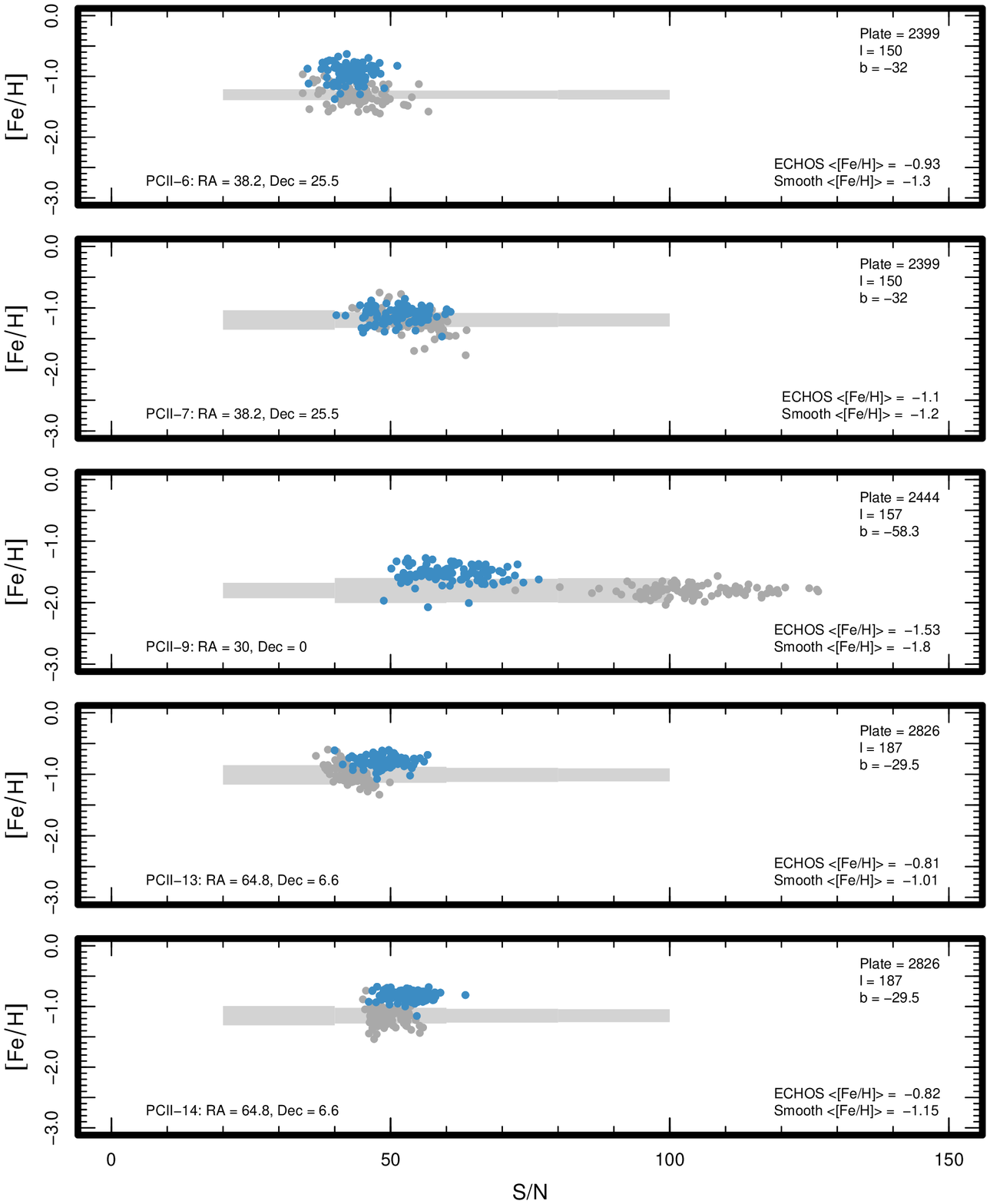}{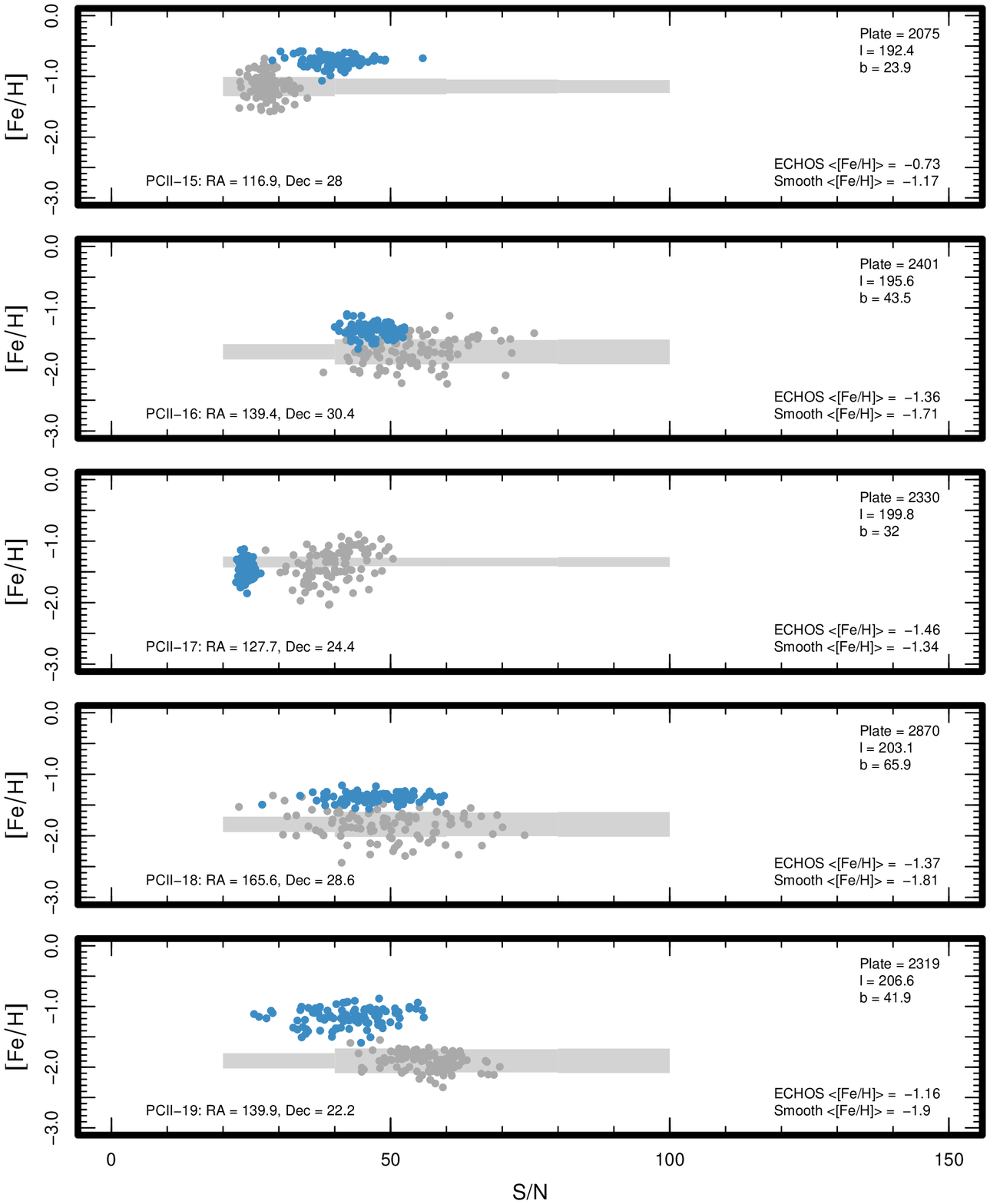}
\caption{\emph{Left}: SSPP [Fe/H] analysis for ECHOS PCII-6, PCII-7,
PCII-9, PCII-13, and PCII-14 from Table~\ref{tbl-3}.  \emph{Right}:
SSPP [Fe/H] analysis for ECHOS PCII-15, PCII-16, PCII-17, PCII-18,
PCII-19 from Table~\ref{tbl-3}.  See the caption to Figure~\ref{fig06}
for a detailed description of this type of figure.\label{fig08}}
\end{figure}

\begin{figure}
\plottwo{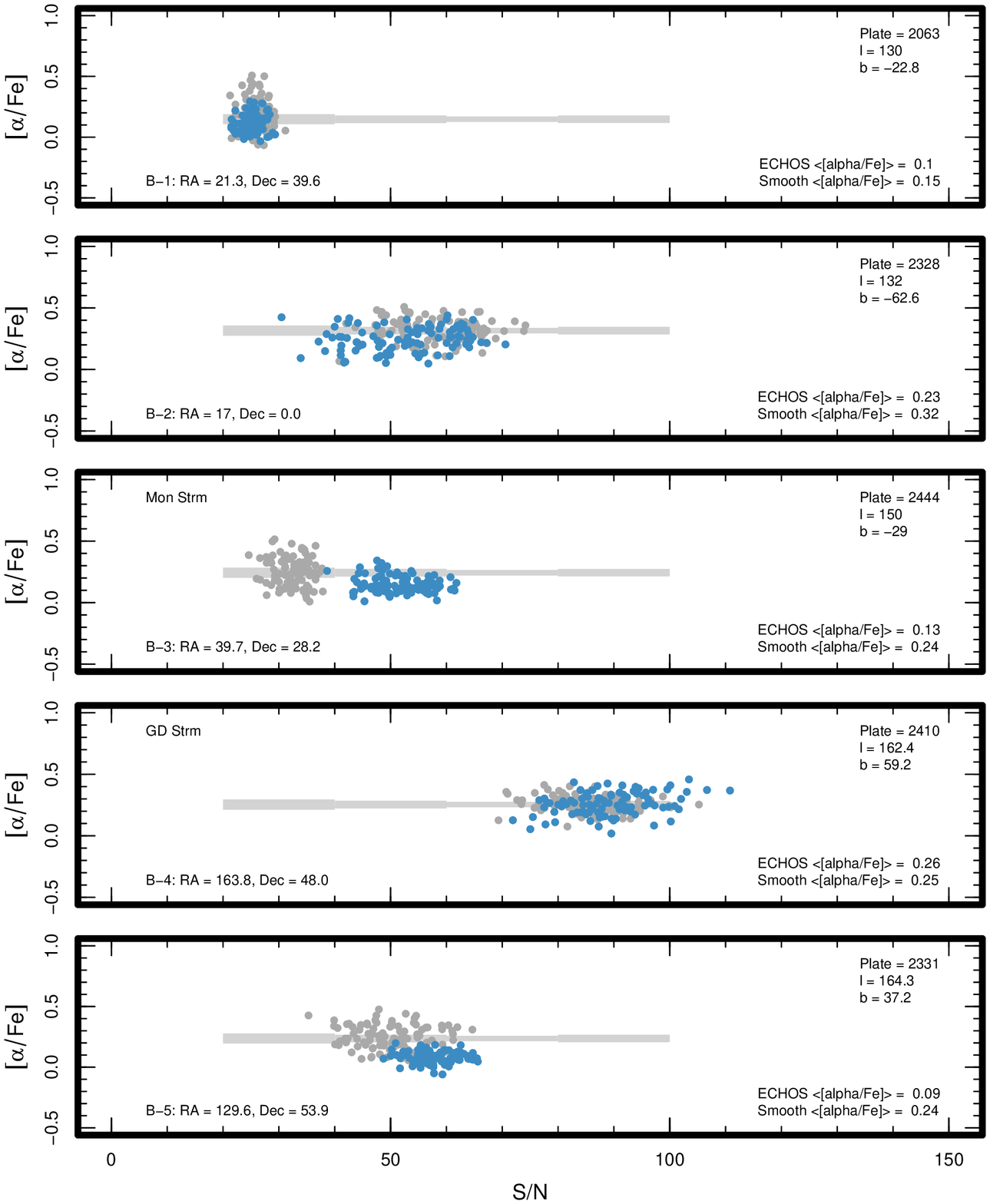}{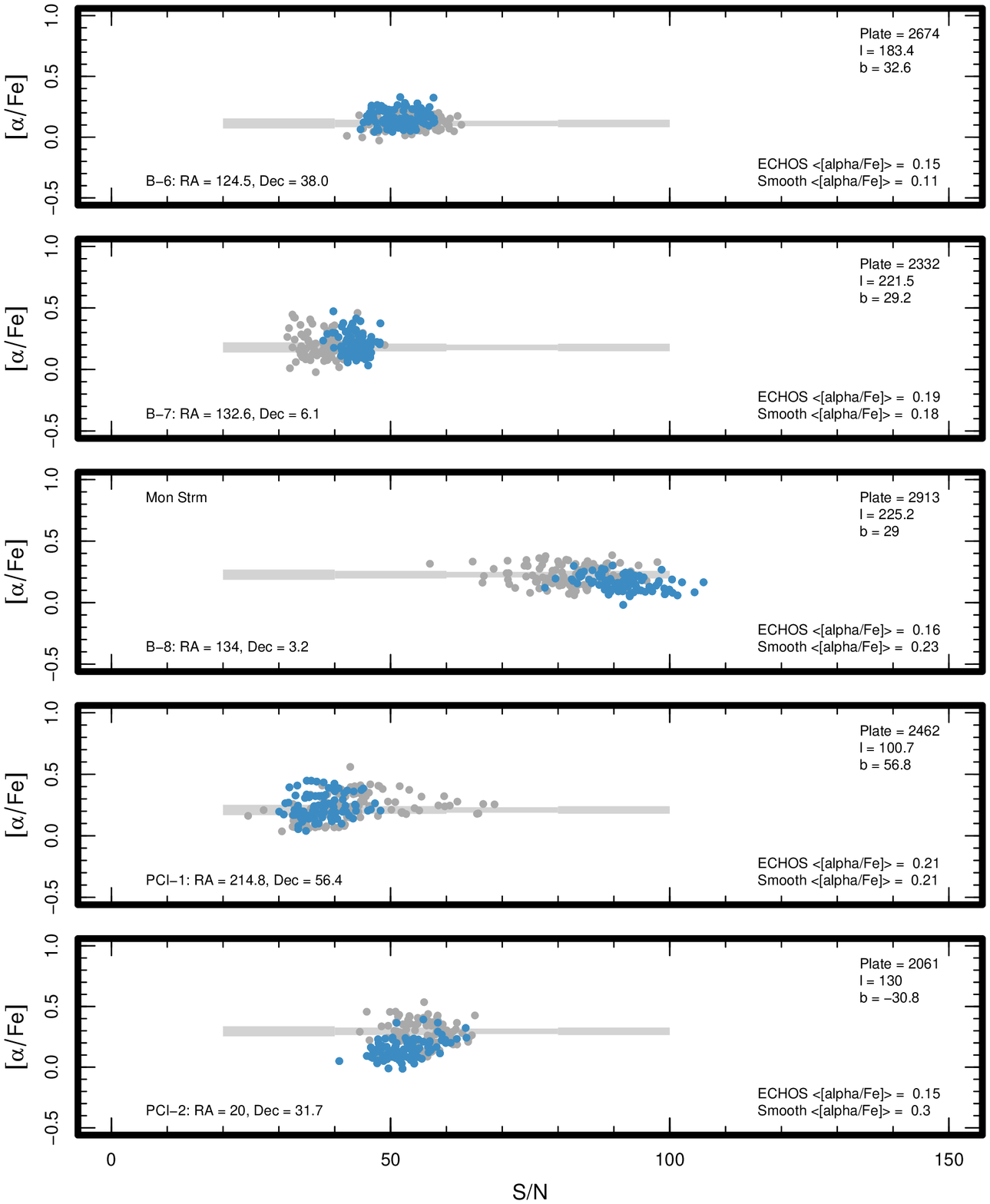}
\caption{SSPP [$\alpha$/Fe] estimates for ECHOS from \citetalias{schl09}
using bootstrap coaddition of SEGUE MPMSTO star spectra.  \emph{Left}:
SSPP [$\alpha$/Fe] analysis for ECHOS B-1, B-2, B-3 (Monoceros),
B-4 \citep{gri06b}, and B-5 from Table~\ref{tbl-1}.  \emph{Right}: SSPP
[$\alpha$/Fe] analysis for ECHOS B-6, B-7, and B-8 from Table~\ref{tbl-1}
as well as PCI-1 and PCI-2 from Table~\ref{tbl-2}.   See the caption
to Figure~\ref{fig06} for a detailed description of this type of
figure.\label{fig09}}
\end{figure}

\begin{figure}
\plottwo{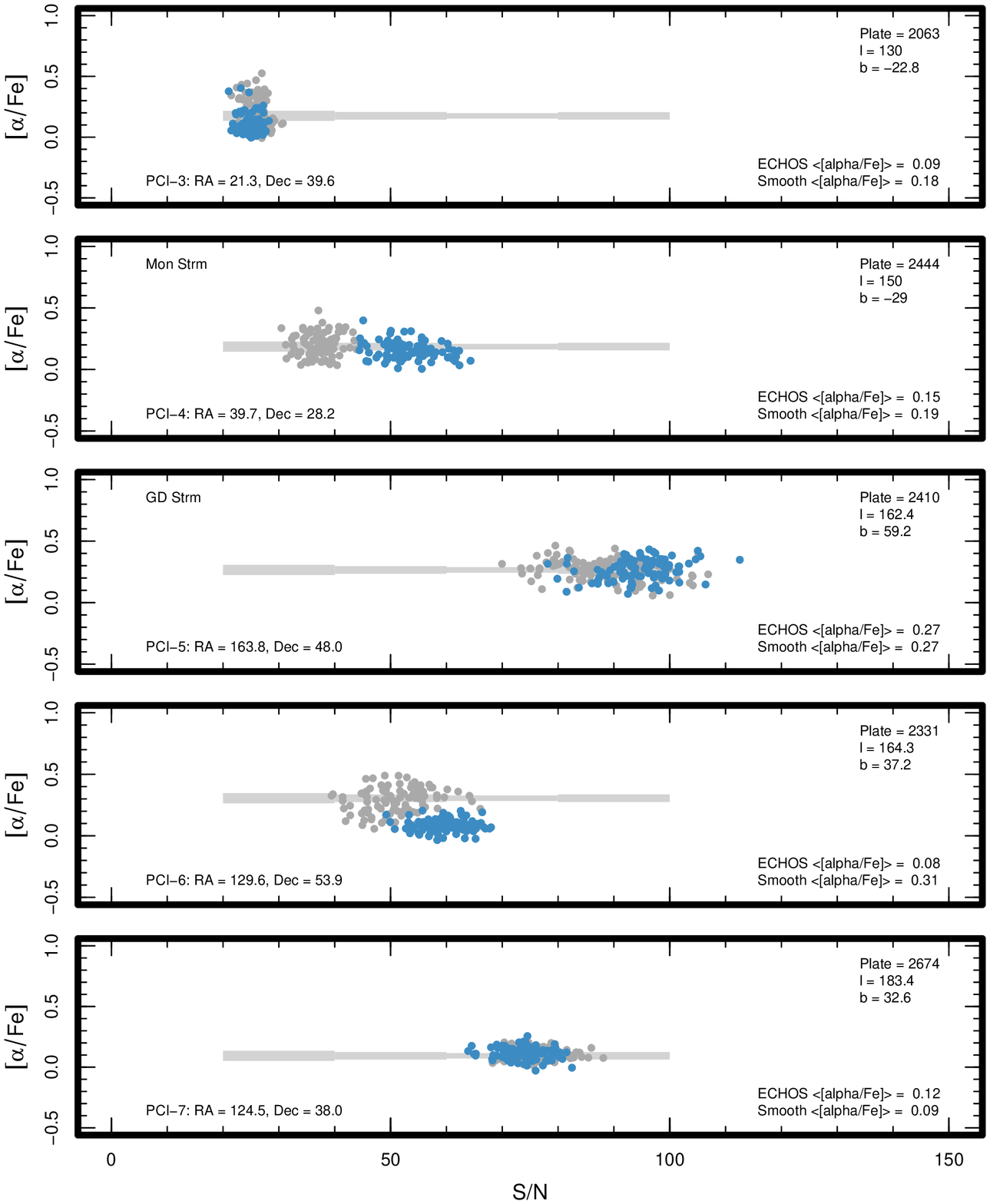}{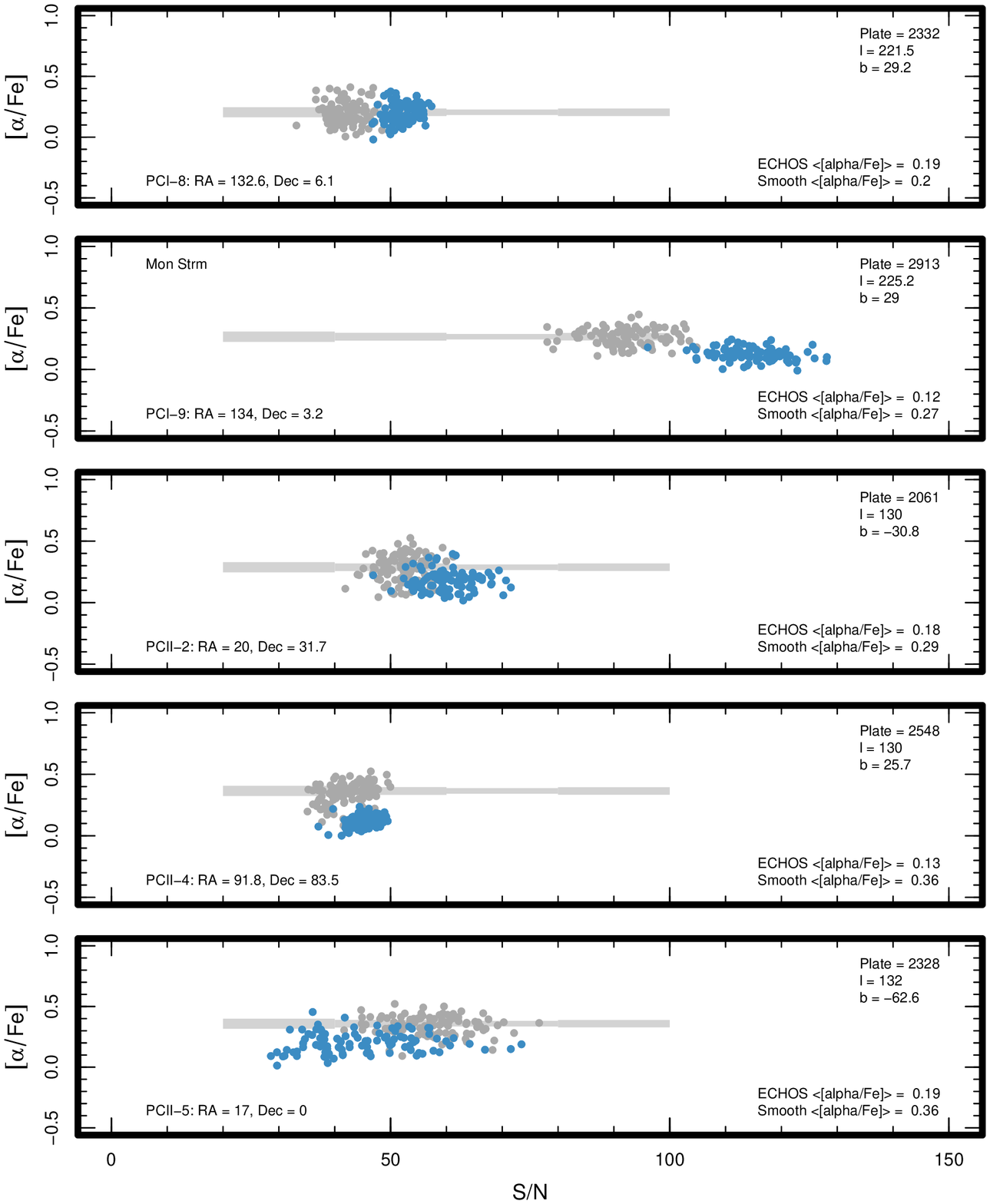}
\caption{\emph{Left}: SSPP [$\alpha$/Fe] analysis for ECHOS PCI-3, PCI-4
(Monoceros), PCI-5 \citep{gri06b}, PCI-6, and PCI-7 from Table~\ref{tbl-2}.
\emph{Right}: SSPP [$\alpha$/Fe] analysis for ECHOS PCI-8 and PCI-9
(Monoceros) from Table~\ref{tbl-2} as well as PCII-2, PCII-4, and PCII-5
from Table~\ref{tbl-3}.  See the caption to Figure~\ref{fig06} for a
detailed description of this type of figure.\label{fig10}}
\end{figure}

\begin{figure}
\plottwo{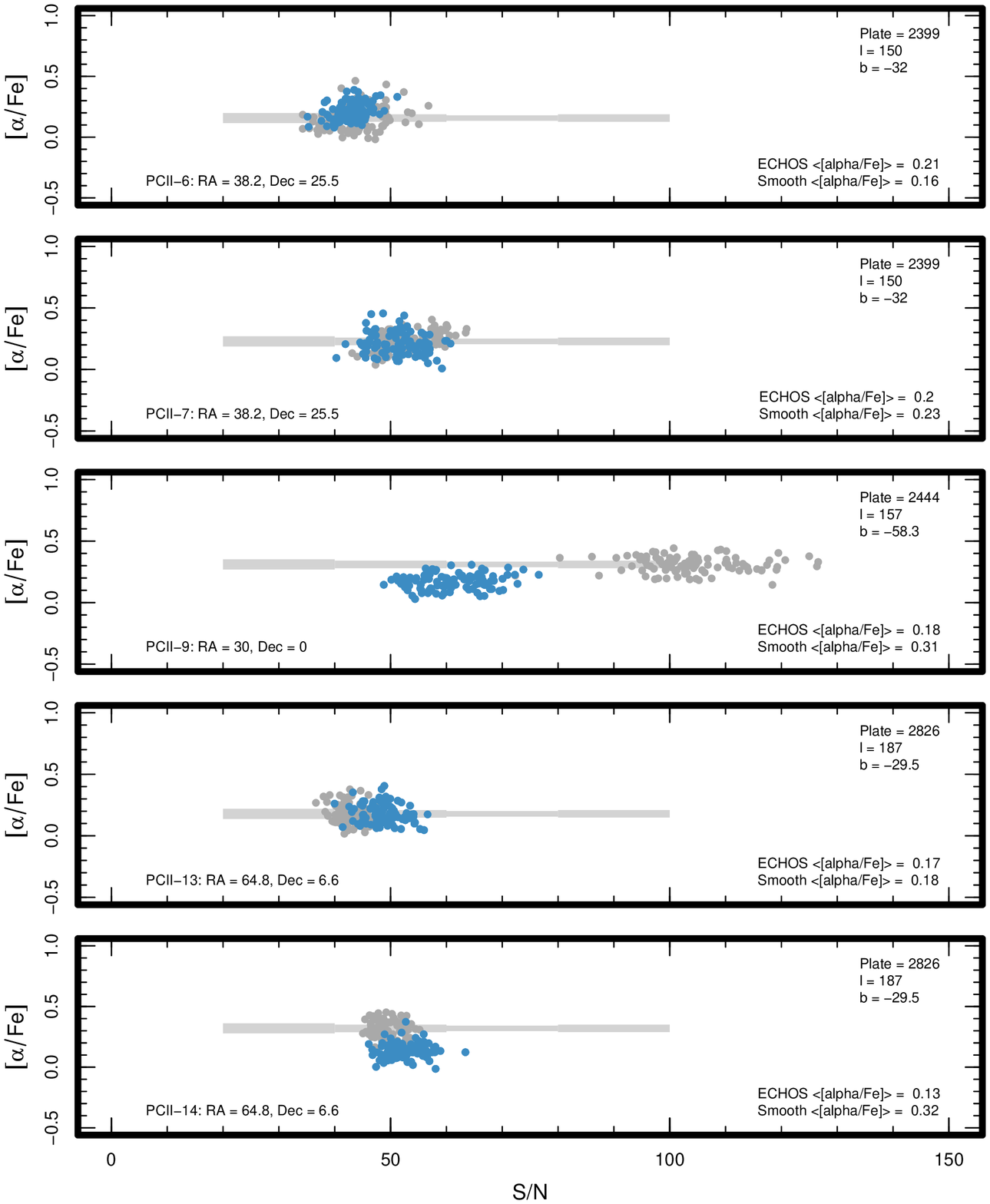}{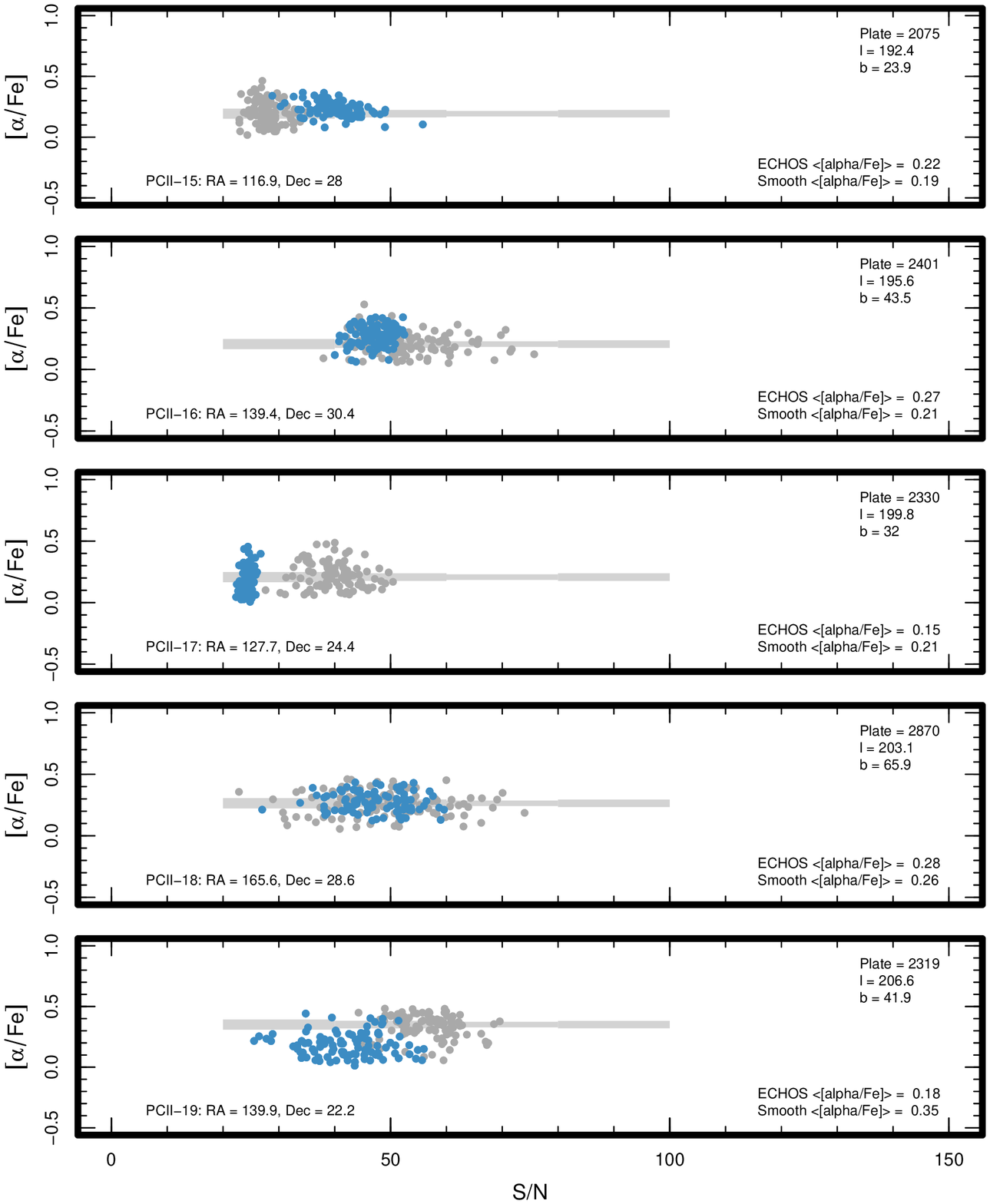}
\caption{\emph{Left}: SSPP [$\alpha$/Fe] analysis for ECHOS PCII-6,
PCII-7, PCII-9, PCII-13, and PCII-14 from Table~\ref{tbl-3}.
\emph{Right}: SSPP [$\alpha$/Fe] analysis for ECHOS PCII-15, PCII-16,
PCII-17, PCII-18, PCII-19 from Table~\ref{tbl-3}.  See the caption
to Figure~\ref{fig06} for a detailed description of this type of
figure.\label{fig11}}
\end{figure}

\begin{figure}
\plottwo{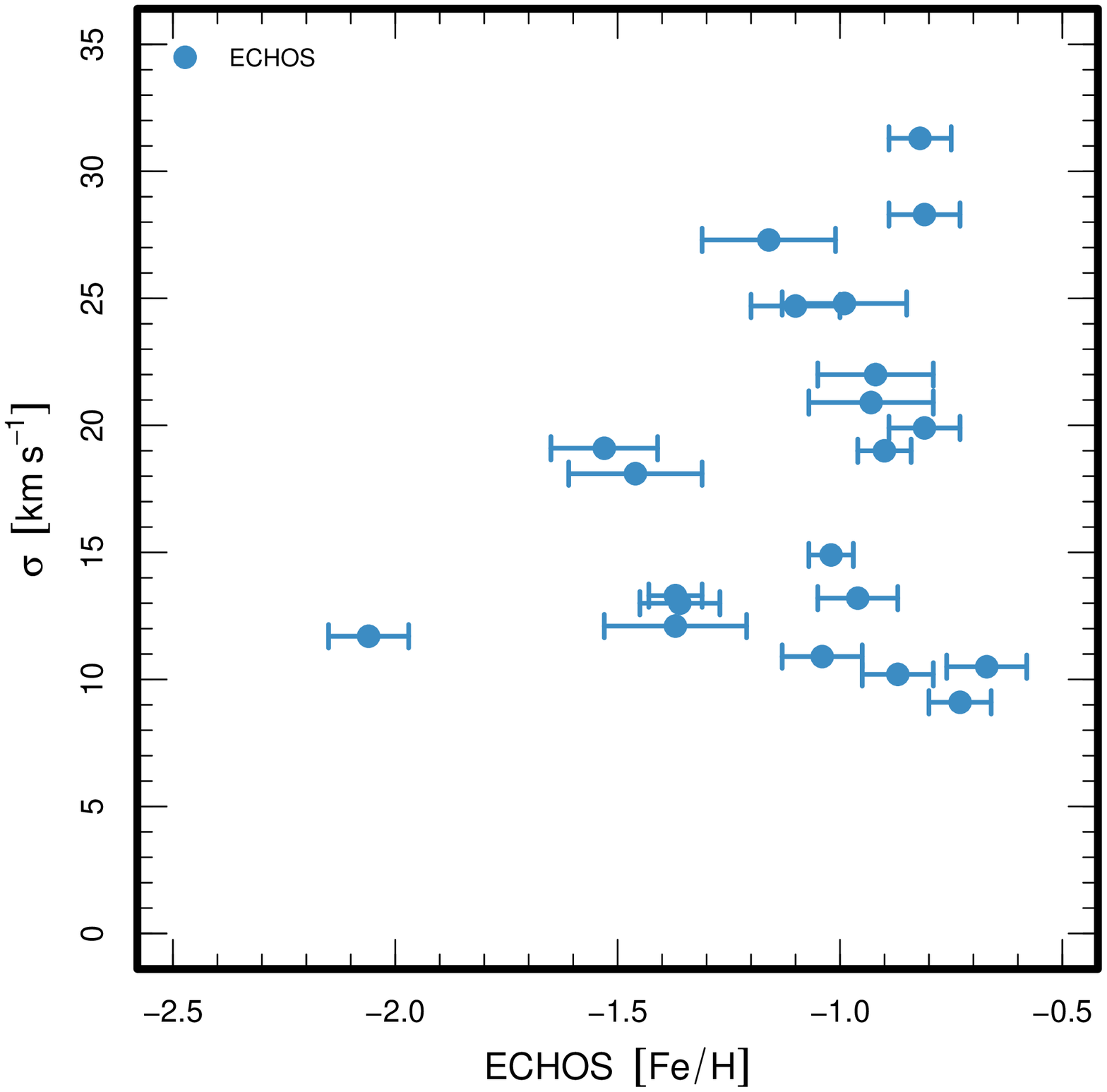}{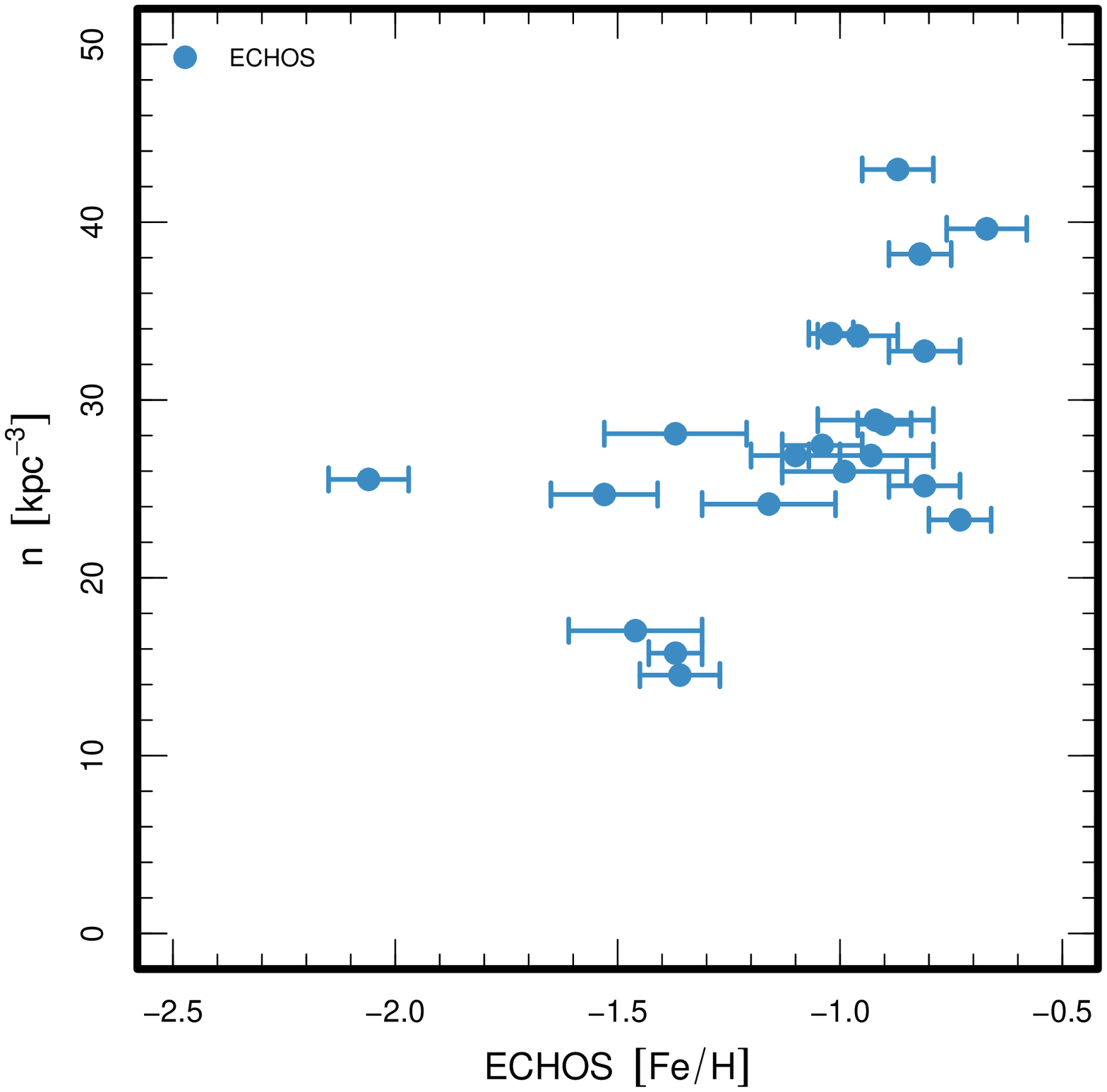}
\caption{Kinematic properties of ECHOS as a function of metallicity.
\emph{Left}: Iron-rich ECHOS can have large velocity dispersions.
\emph{Right}: Iron-rich ECHOS have the highest space densities;
this prominence of metal-rich substructures was predicted in
\citet{fon06}.\label{fig12}}
\end{figure}

\begin{figure}
\plottwo{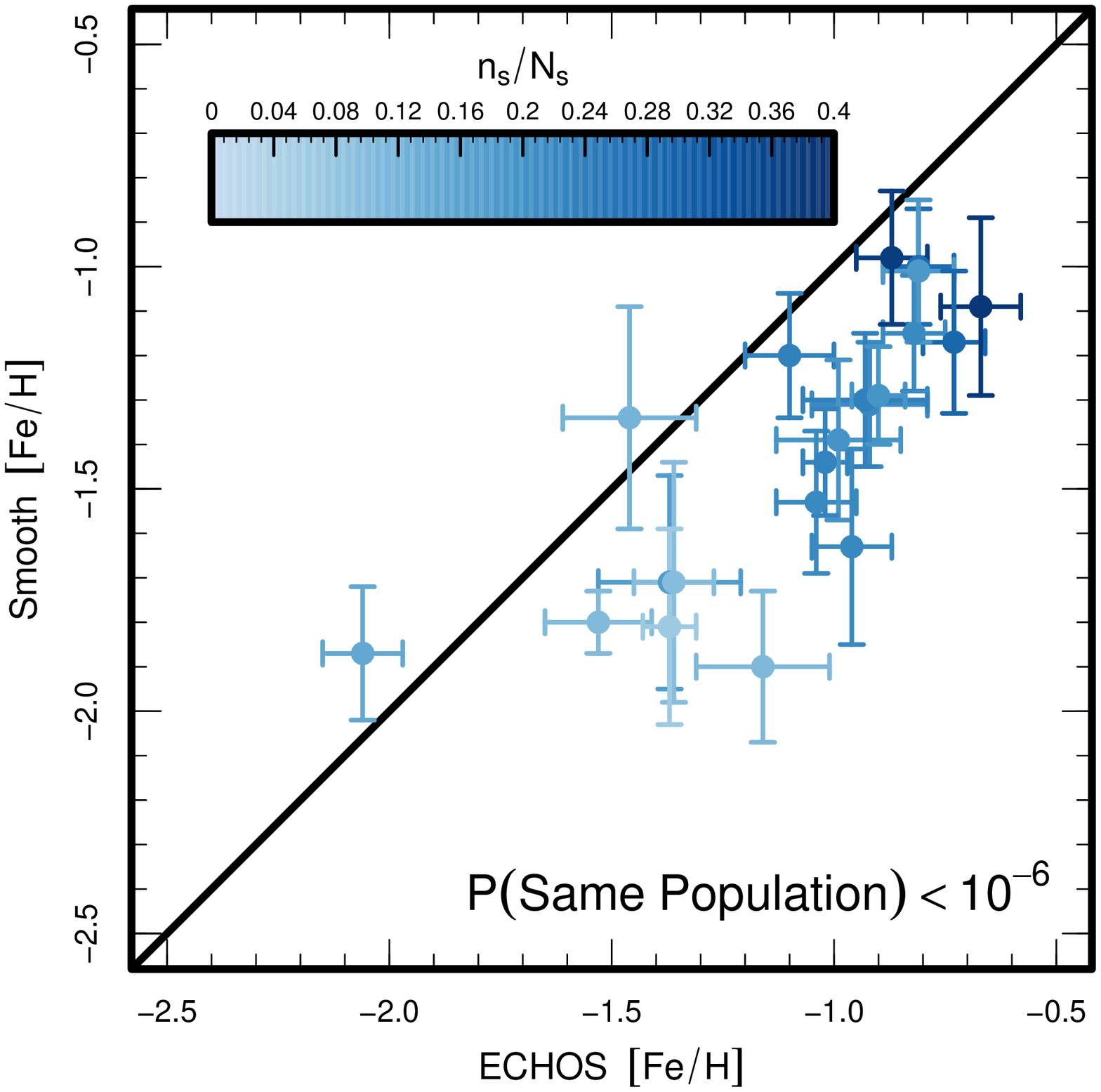}{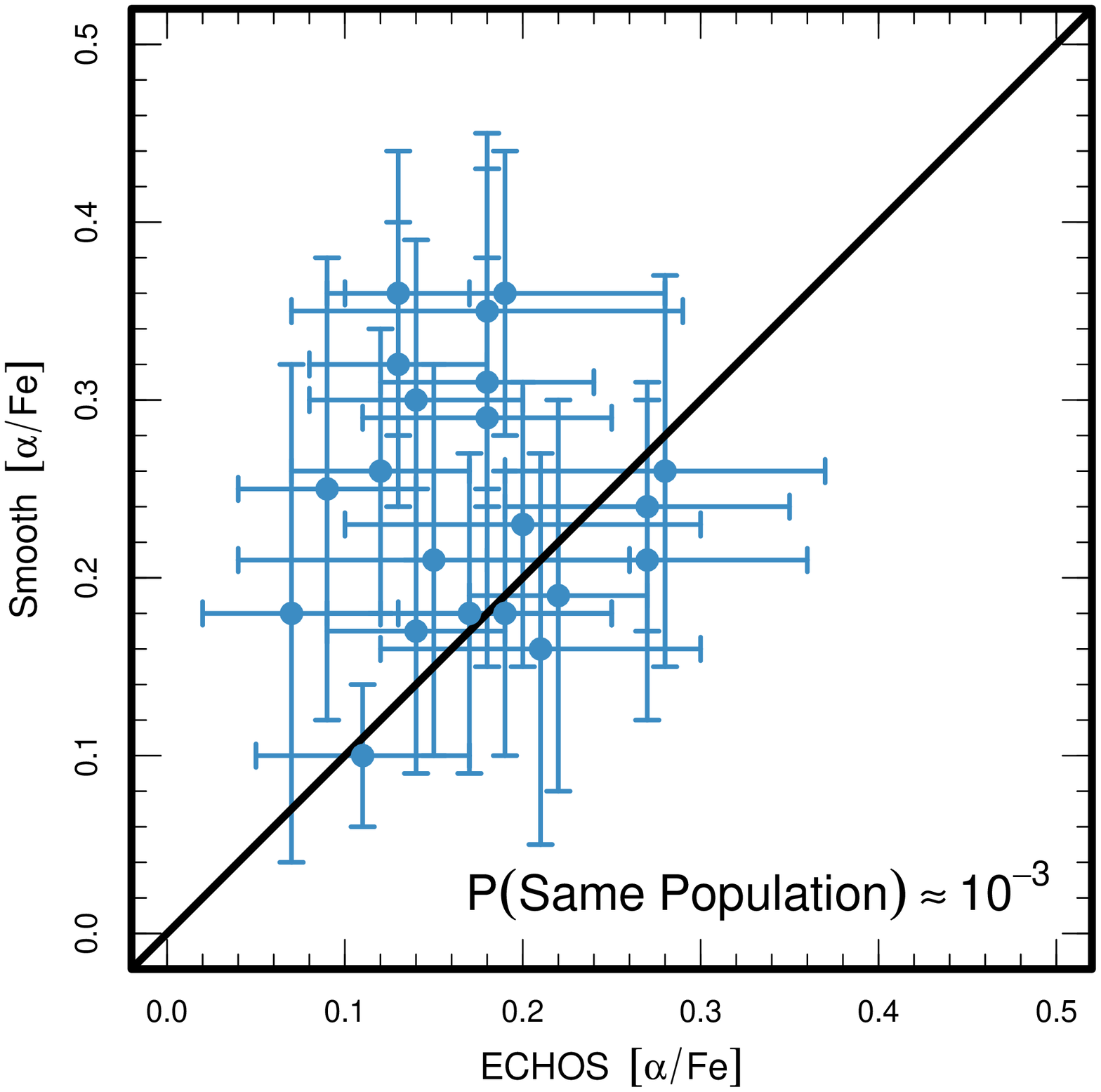}
\caption{Comparison of the metallicity of ECHOS to the metallicity
of the MPMSTO population in the kinematically smooth component of the
inner halo along the same line of sight where the ECHOS was discovered.
\emph{Left}: ECHOS are more iron-rich than the MPMSTO stars associated
with the kinematically smooth inner halo.  Moreover, ECHOS are so
iron-rich that an origin within one or more now tidally disrupted
ultrafaint dwarf galaxies is unlikely.  That is, the mass--luminosity
relation of Milky Way dwarf spheroidal galaxies \citep[e.g.,][]{kir08b}
requires a progenitor luminosity $L \sim 10^{8}~L_{\odot}$ to reach a mean
iron metallicity [Fe/H] $\sim -1.0$.  Lines of sight hosting an ECHOS
for which we infer a smooth component metallicity [Fe/H] $\gtrsim -1.6$
are those lines of sight along which the ratio of the number of MPMSTO
stars with radial velocities consistent with ECHOS membership $n_s$ to
the total number of MPMSTO stars with radial velocity measurements $N_s$
is non-negligible, or $n_s/N_s \gtrsim 0$.  In other words, lines of sight
where ECHOS are a significant contributor to the MPMSTO population tend
to have inferred smooth component metallicities [Fe/H] $\gtrsim -1.6$.
As a result, those apparently enhanced iron metallicities in the smooth
component are likely due to the presence of stars that are outside the
radial velocity overdensity that defines the ECHOS, but that are still
associated with ECHOS.  \emph{Right}: ECHOS are less $\alpha$-enhanced
than the MPMSTO stars associated with the kinematically smooth inner halo,
so (modulo any changes in the initial mass function) the MPMSTO stars in
ECHOS formed in environments in which the star formation timescale was
long relative to the star formation timescale in the massive progenitor
of the bulk of the inner halo \citep[e.g.,][]{rob05}.\label{fig13}}
\end{figure}

\begin{figure}
\plotone{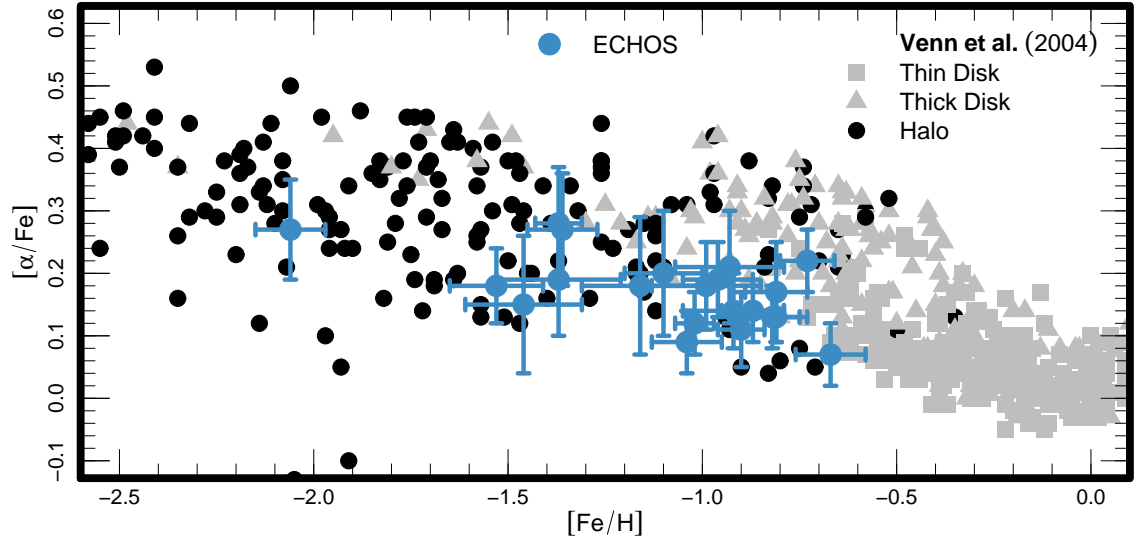}
\caption{Distribution of ECHOS in the [Fe/H]--[$\alpha$/Fe] plane
along with individual stars collected from the literature as presented
in \citet{ven04}.  The ECHOS mostly fall in a region of that plane
that is relatively sparsely occupied by--but not completely barren
of--individual stars.  In general, the ECHOS are more iron-poor than
thick-disk stars and more iron-poor and $\alpha$-enhanced than typical
thin-disk stars.\label{fig14}}
\end{figure}

\begin{figure}
\plotone{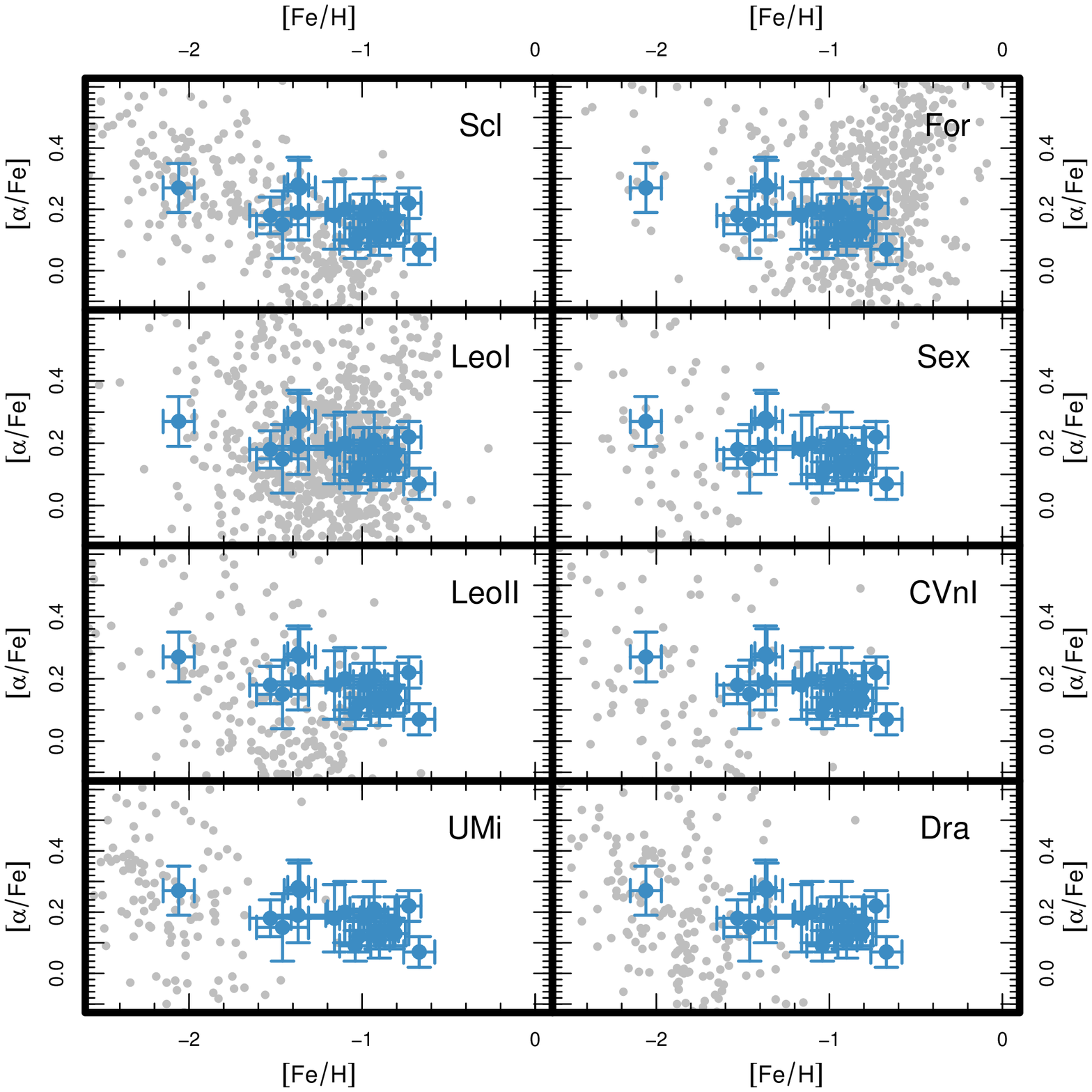}
\caption{Distribution of ECHOS in the [Fe/H]--[$\alpha$/Fe] plane
along with individual stars in dSph galaxies from  \citet{kir10}.
In each panel, the gray points in the background represent [Fe/H] and
[$\alpha$/Fe] measurements for individual dSph stars.  The blue points
in the foreground are our [Fe/H] and [$\alpha$/Fe] measurements for
individual ECHOS.  Left to right and top to bottom, the dSph galaxies are:
Sculptor, Fornax, Leo I, Sextans, Leo II, Canes Venatici I, Ursa Minor,
and Draco.  ECHOS are plausibly associated with a progenitor comparable
to Sculptor or Leo I.\label{fig15}}
\end{figure}

\begin{figure}
\plotone{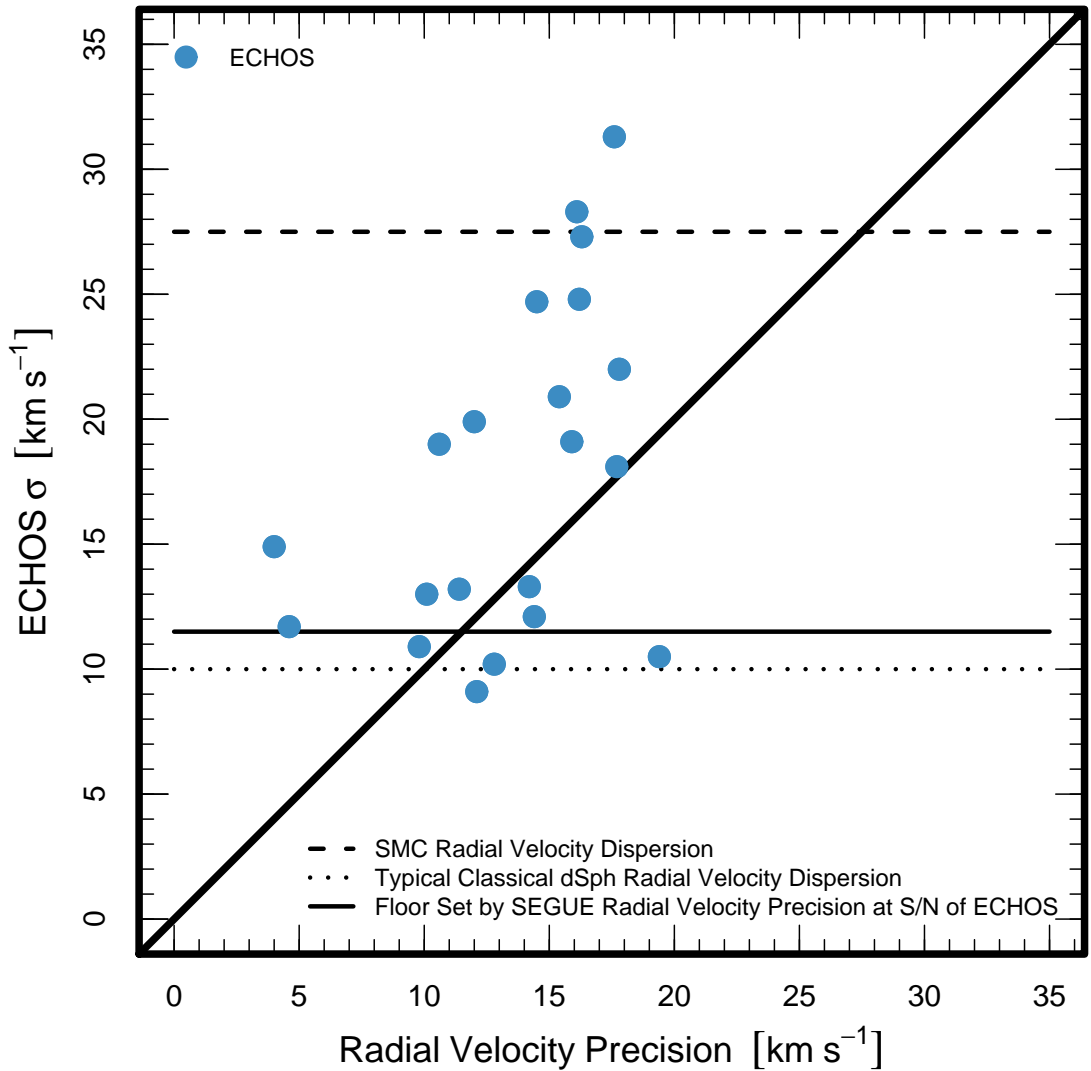}
\caption{Velocity dispersion of ECHOS as a function of the median radial
velocity precision of the stars in the ECHOS.  The fact that the points
are above the line $y=x$ suggests that the observed velocity dispersion
is not only a function of the limited radial velocity precision for
faint MPMSTO stars.  The velocity dispersion observed in an ECHOS is
a lower limit on the velocity dispersion of any gravitationally bound
progenitor.  This results from the fact that the phase-space distribution
of a disrupted stellar system becomes colder with time, combined with the
fact that each SEGUE line of sight would only intersect a small fraction
of an orbit of an element of halo substructure.  Typical classical dwarf
spheroidal galaxies have radial velocity dispersions $\sigma \sim 10$ km
s$^{-1}$ \citep[e.g.,][]{mat98}, while the SMC has a velocity dispersion
of $\sigma \approx 27.5 \pm 0.5$ km s$^{-1}$ \citep[e.g.,][]{har06}.
The origin of ECHOS through the gravitational disruption of a progenitor
with a velocity dispersion $\sigma > 20$ km s$^{-1}$ implies a progenitor
mass $M_{\mathrm{tot}} \gtrsim 10^{9}~M_{\odot}$.\label{fig16}}
\end{figure}

A single iteration of these eight steps produces a single estimate of
the mean metallicity of the input sample, corresponding to a single
point in Figures~\ref{fig01} through~\ref{fig04} and Figures~\ref{fig06}
through~\ref{fig11}.  We repeat these eight steps many times to produce
many estimates of the mean metallicity of the input sample.  We use that
distribution to estimate both the true mean metallicity of the input
sample and our uncertainty in that estimate.

\clearpage

\clearpage

\clearpage

\clearpage

\clearpage

\clearpage

\clearpage

\clearpage

\clearpage

\clearpage

\clearpage

\clearpage

\clearpage

\clearpage

\clearpage

\clearpage

\clearpage
\begin{deluxetable}{lrrrrrrrrrrrrrrrrrrrrrc}
\tablecaption{Properties of ECHOS\label{tbl-1}}
\tablewidth{0pt}
\tablehead{\colhead{ID} & \colhead{RA} & \colhead{Dec} & \colhead{l} &
\colhead{b} & \colhead{bplate} & \colhead{fplate} & \colhead{N$_s$} &
\colhead{N$_p$} & \colhead{Vol} & \colhead{d} & \colhead{err} &
\colhead{v$_r$} & \colhead{n$_s$} & \colhead{[Fe/H]$_E$} & \colhead{err} &
\colhead{[Fe/H]$_S$} & \colhead{err} & \colhead{[$\alpha$/Fe]$_E$} &
\colhead{err} & \colhead{[$\alpha$/Fe]$_S$} & \colhead{err}\\
\colhead{} & \colhead{(deg)} & \colhead{(deg)} & \colhead{(deg)} &
\colhead{(deg)} & \colhead{} & \colhead{} & \colhead{} & \colhead{} &
\colhead{(kpc$^3$)} & \colhead{(kpc)} & \colhead{(kpc)} &
\colhead{(km s$^{-1}$)} & \colhead{} & \colhead{(dex)} & \colhead{(dex)} &
\colhead{(dex)} & \colhead{(dex)} & \colhead{(dex)} & \colhead{(dex)} &
\colhead{(dex)} & \colhead{(dex)}}
\startdata
B-1 &  21.3 & 39.6 & 130.0 & -22.8 & 2043 & 2063 &  34 & 228 & 2.20 & 18.4 & 3.2 & -130 & 12 & -0.70 & 0.12 & -0.99 & 0.18 & 0.10 & 0.07 & 0.15 & 0.13 \\
B-2 &  17.0 &  0.0 & 132.0 & -62.6 & 2313 & 2328 & 109 & 561 & 3.48 & 10.8 & 6.6 & -170 & 20 & -1.33 & 0.14 & -1.71 & 0.25 & 0.23 & 0.10 & 0.32 & 0.10 \\
B-3\tablenotemark{a} &  39.7 & 28.2 & 150.0 & -29.0 & 2442 & 2444 &  59 & 265 & 2.30 & 11.2 & 2.0 &  -50 & 17 & -0.87 & 0.09 & -0.90 & 0.15 & 0.13 & 0.06 & 0.24 & 0.12 \\
B-4\tablenotemark{b} & 163.8 & 48.0 & 162.4 &  59.2 & 2390 & 2410 & 150 & 672 & 3.86 &  6.6 & 2.8 & -130 & 25 & -2.12 & 0.11 & -1.82 & 0.13 & 0.26 & 0.10 & 0.25 & 0.07 \\
B-5 & 129.6 & 53.9 & 164.3 &  37.2 & 2316 & 2331 &  93 & 425 & 3.33 &  9.6 & 2.5 &  -10 & 20 & -1.02 & 0.09 & -1.51 & 0.17 & 0.09 & 0.04 & 0.24 & 0.12 \\
B-6 & 124.5 & 38.0 & 183.4 &  32.6 & 2670 & 2674 &  83 & 514 & 3.46 &  9.9 & 1.1 &   30 & 17 & -0.97 & 0.06 & -1.10 & 0.13 & 0.15 & 0.07 & 0.11 & 0.06 \\
B-7 & 132.6 &  6.1 & 221.5 &  29.2 & 2317 & 2332 &  69 & 470 & 3.04 & 11.0 & 1.3 &   70 & 17 & -0.93 & 0.10 & -1.50 & 0.18 & 0.19 & 0.08 & 0.18 & 0.09 \\
B-8\tablenotemark{a} & 134.0 &  3.2 & 225.2 &  29.0 & 2888 & 2913 &  74 & 514 & 3.50 & 11.0 & 1.7 &   90 & 19 & -1.08 & 0.05 & -1.33 & 0.14 & 0.16 & 0.07 & 0.23 & 0.10
\enddata
\tablecomments{Kinematic and metallicity data for all bin detection ECHOS
from \citetalias{schl09}.  The columns are: right ascension, declination,
galactic longitude, galactic latitude, SEGUE bright plate number, SEGUE
faint plate number, number of SEGUE MPMSTO spectra along the indicated
line of sight, number of photometric MPMSTO candidates along the indicated
line of sight, volume scanned along the indicated line of sight, median
distance of the stars in the ECHOS, the error in that estimate, the radial
velocity of the ECHOS, the number of MPMSTO stars kinematically associated
with the ECHOS, the iron metallicity of the ECHOS, the error in that
estimate, the iron metallicity of the kinematically smooth component of
the halo along the indicated line of sight, the error in that estimate,
the $\alpha$-enhancement of the ECHOS, the error in that estimate, the
$\alpha$-enhancement of the kinematically smooth component of the halo
along the indicated line of sight, and the error in that estimate.}
\tablenotetext{a}{Monoceros Stream}
\tablenotetext{b}{\citet{gri06b} Stream}
\end{deluxetable}

\clearpage
\begin{deluxetable}{lrrrrrrrrrrrrrrrrrrrrrrrc}
\tablecaption{Properties of ECHOS\label{tbl-2}}
\tablewidth{0pt}
\tablehead{\colhead{ID} & \colhead{RA} & \colhead{Dec} & \colhead{l} &
\colhead{b} & \colhead{bplate} & \colhead{fplate} & \colhead{N$_s$} &
\colhead{N$_p$} & \colhead{Vol} & \colhead{d} & \colhead{err} &
\colhead{v$_r$} & \colhead{$\sigma$} & \colhead{err} & \colhead{n$_s$} &
\colhead{[Fe/H]$_E$} & \colhead{err} & \colhead{[Fe/H]$_S$} & \colhead{err} &
\colhead{[$\alpha$/Fe]$_E$} & \colhead{err} & \colhead{[$\alpha$/Fe]$_S$} &
\colhead{err}\\
\colhead{} & \colhead{(deg)} & \colhead{(deg)} & \colhead{(deg)} &
\colhead{(deg)} & \colhead{} & \colhead{} & \colhead{} &
\colhead{} & \colhead{(kpc$^{3}$)} & \colhead{(kpc)} & \colhead{(kpc)} &
\colhead{(km s$^{-1}$)} & \colhead{(km s$^{-1}$)} & \colhead{(km s$^{-1}$)} &
\colhead{} & \colhead{(dex)} & \colhead{(dex)} & \colhead{(dex)} &
\colhead{(dex)} & \colhead{(dex)} & \colhead{(dex)} & \colhead{(dex)} &
\colhead{(dex)}}
\startdata
PCI-1 & 214.8 & 56.4 & 100.7 &  56.8 & 2447 & 2462 & 122 & 673 & 2.79 &  8.7 & 2.3 & -328 & 15.1 & 11.5 &  8 & -1.73 & 0.24 & -1.73 & 0.22 & 0.21 & 0.07 & 0.21 & 0.15 \\
PCI-2 &  20.0 & 31.7 & 130.0 & -30.8 & 2041 & 2061 &  93 & 349 & 2.60 & 13.4 & 3.3 & -125 & 22.0 & 17.8 & 20 & -0.95 & 0.12 & -1.28 & 0.15 & 0.15 & 0.08 & 0.30 & 0.10 \\
PCI-3 &  21.3 & 39.6 & 130.0 & -22.8 & 2043 & 2063 &  34 & 228 & 2.20 & 18.7 & 3.5 & -121 & 10.5 & 19.4 & 13 & -0.65 & 0.12 & -1.11 & 0.16 & 0.09 & 0.06 & 0.18 & 0.14 \\
PCI-4\tablenotemark{a} &  39.7 & 28.2 & 150.0 & -29.0 & 2442 & 2444 &  59 & 265 & 2.30 & 11.4 & 2.6 &  -57 & 10.2 & 12.8 & 22 & -0.87 & 0.07 & -0.98 & 0.19 & 0.15 & 0.07 & 0.19 & 0.09 \\
PCI-5\tablenotemark{b} & 163.8 & 48.0 & 162.4 &  59.2 & 2930 & 2410 & 150 & 672 & 3.86 &  6.6 & 2.5 & -132 & 11.7 &  4.6 & 22 & -2.06 & 0.09 & -1.85 & 0.15 & 0.27 & 0.09 & 0.27 & 0.09 \\
PCI-6 & 129.6 & 53.9 & 164.3 &  37.2 & 2316 & 2331 &  93 & 425 & 3.33 & 10.0 & 2.4 &  -13 & 10.9 &  9.8 & 20 & -1.03 & 0.09 & -1.51 & 0.21 & 0.08 & 0.04 & 0.31 & 0.12 \\
PCI-7 & 124.5 & 38.0 & 183.4 &  32.6 & 2670 & 2674 &  83 & 514 & 3.46 & 10.1 & 1.5 &   29 & 19.0 & 10.6 & 16 & -0.91 & 0.06 & -1.29 & 0.12 & 0.12 & 0.06 & 0.09 & 0.05 \\
PCI-8 & 132.6 &  6.1 & 221.5 &  29.2 & 2317 & 2332 &  69 & 470 & 3.04 & 10.9 & 1.6 &   71 & 13.2 & 11.4 & 15 & -0.95 & 0.09 & -1.61 & 0.15 & 0.19 & 0.09 & 0.20 & 0.10 \\
PCI-9\tablenotemark{a} & 134.0 &  3.2 & 225.2 &  29.0 & 2888 & 2913 &  74 & 514 & 3.50 & 10.3 & 1.4 &   85 & 14.9 &  4.0 & 17 & -1.01 & 0.05 & -1.44 & 0.15 & 0.12 & 0.05 & 0.27 & 0.08
\enddata
\tablecomments{Kinematic and metallicity data for all class I peak
detection ECHOS from \citetalias{schl09}.  The columns are: right
ascension, declination, galactic longitude, galactic latitude, SEGUE
bright plate number, SEGUE faint plate number, number of SEGUE MPMSTO
spectra along the indicated line of sight, number of photometric MPMSTO
candidates along the indicated line of sight, volume scanned along the
indicated line of sight, median distance of the stars in the ECHOS, the
error in that estimate, the radial velocity of the ECHOS, the velocity
dispersion of the ECHOS, the median radial velocity error of the stars in
the ECHOS, the number of MPMSTO stars kinematically associated with the
ECHOS, the iron metallicity of the ECHOS, the error in that estimate,
the iron metallicity of the kinematically smooth component of the
halo along the indicated line of sight, the error in that estimate,
the $\alpha$-enhancement of the ECHOS, the error in that estimate, the
$\alpha$-enhancement of the kinematically smooth component of the halo
along the indicated line of sight, and the error in that estimate.}
\tablenotetext{a}{Monoceros Stream}
\tablenotetext{b}{\citet{gri06b} Stream}
\end{deluxetable}

\clearpage
\begin{deluxetable}{lrrrrrrrrrrrrrrrrrrrrrrrc}
\tablecaption{Properties of ECHOS\label{tbl-3}}
\tablewidth{0pt}
\tablehead{\colhead{ID} & \colhead{RA} & \colhead{Dec} & \colhead{l} &
\colhead{b} & \colhead{bplate} & \colhead{fplate} & \colhead{N$_s$} &
\colhead{N$_p$} & \colhead{Vol} & \colhead{d} & \colhead{err} &
\colhead{v$_r$} & \colhead{$\sigma$} & \colhead{err} & \colhead{n$_s$} &
\colhead{[Fe/H]$_E$} & \colhead{err} & \colhead{[Fe/H]$_S$} & \colhead{err} &
\colhead{[$\alpha$/Fe]$_E$} & \colhead{err} & \colhead{[$\alpha$/Fe]$_S$} &
\colhead{err}\\
\colhead{} & \colhead{(deg)} & \colhead{(deg)} & \colhead{(deg)} &
\colhead{(deg)} & \colhead{} & \colhead{} & \colhead{} &
\colhead{} & \colhead{(kpc$^{3}$)} & \colhead{(kpc)} & \colhead{(kpc)} &
\colhead{(km s$^{-1}$)} & \colhead{(km s$^{-1}$)} & \colhead{(km s$^{-1}$)} &
\colhead{} & \colhead{(dex)} & \colhead{(dex)} & \colhead{(dex)} &
\colhead{(dex)} & \colhead{(dex)} & \colhead{(dex)} & \colhead{(dex)} &
\colhead{(dex)}}
\startdata
PCII-1 &  20.0 & 31.7 & 130.0 & -30.8 & 2041 & 2061 &  93 & 349 & 2.60 & 13.4 & 3.3 & -125 & 22.0 & 17.8 & 20 & -0.92 & 0.13 & -1.31 & 0.14 & 0.14 & 0.06 & 0.30 & 0.09 \\
PCII-2 &  20.0 & 31.7 & 130.0 & -30.8 & 2041 & 2061 &  93 & 349 & 2.60 & 12.7 & 3.4 &  -98 & 24.8 & 16.2 & 18 & -0.99 & 0.14 & -1.39 & 0.18 & 0.18 & 0.07 & 0.29 & 0.14 \\
PCII-3 &  21.3 & 39.6 & 130.0 & -22.8 & 2043 & 2063 &  34 & 228 & 2.20 & 18.7 & 3.5 & -121 & 10.5 & 19.4 & 13 & -0.67 & 0.09 & -1.09 & 0.20 & 0.07 & 0.05 & 0.18 & 0.14 \\
PCII-4 &  91.8 & 83.5 & 130.0 &  25.7 & 2540 & 2548 &  47 & 223 & 2.45 & 12.7 & 1.9 &  -95 & 19.9 & 12.0 & 13 & -0.81 & 0.08 & -1.00 & 0.13 & 0.13 & 0.04 & 0.36 & 0.08 \\
PCII-5 &  17.0 &  0.0 & 132.0 & -62.6 & 2313 & 2328 & 109 & 561 & 3.48 & 12.0 & 6.1 & -173 & 12.1 & 14.4 & 19 & -1.37 & 0.16 & -1.71 & 0.24 & 0.19 & 0.09 & 0.36 & 0.08 \\
PCII-6 &  38.2 & 25.5 & 150.0 & -32.0 & 2379 & 2399 &  60 & 273 & 2.37 & 11.6 & 2.7 &  -93 & 20.9 & 15.4 & 14 & -0.93 & 0.14 & -1.30 & 0.15 & 0.21 & 0.09 & 0.16 & 0.11 \\
PCII-7 &  38.2 & 25.5 & 150.0 & -32.0 & 2379 & 2399 &  60 & 273 & 2.37 & 11.0 & 2.3 &  -66 & 24.7 & 14.5 & 14 & -1.10 & 0.10 & -1.20 & 0.14 & 0.20 & 0.10 & 0.23 & 0.08 \\
PCII-8\tablenotemark{a} &  39.7 & 28.2 & 150.0 & -29.0 & 2045 & 2065 &  59 & 265 & 2.30 & 13.9 & 6.8 &  -57 & 10.2 & 12.8 & 22 & -0.87 & 0.08 & -0.98 & 0.15 & 0.14 & 0.05 & 0.17 & 0.08 \\
PCII-9 &  30.0 &  0.0 & 157.0 & -58.3 & 2442 & 2444 & 173 & 987 & 4.16 & 10.1 & 2.4 & -177 & 19.1 & 15.9 & 18 & -1.53 & 0.12 & -1.80 & 0.07 & 0.18 & 0.06 & 0.31 & 0.07 \\
PCII-10\tablenotemark{b} & 163.8 & 48.0 & 162.4 &  59.2 & 2390 & 2410 & 150 & 672 & 3.86 &  6.6 & 2.5 & -132 & 11.7 &  4.6 & 22 & -2.06 & 0.09 & -1.87 & 0.15 & 0.27 & 0.08 & 0.24 & 0.07 \\
PCII-11 & 129.6 & 53.9 & 164.3 &  37.2 & 2316 & 2331 &  93 & 425 & 3.33 & 10.0 & 2.4 &  -13 & 10.9 &  9.8 & 20 & -1.04 & 0.09 & -1.53 & 0.16 & 0.09 & 0.05 & 0.25 & 0.13 \\
PCII-12 & 124.5 & 38.0 & 183.4 &  32.6 & 2670 & 2674 &  83 & 514 & 3.46 & 10.1 & 1.5 &   29 & 19.0 & 10.6 & 16 & -0.90 & 0.06 & -1.29 & 0.11 & 0.11 & 0.06 & 0.10 & 0.04 \\
PCII-13 &  64.8 &  6.6 & 187.0 & -29.5 & 2805 & 2826 &  65 & 353 & 1.99 & 11.9 & 1.9 &   20 & 28.3 & 16.1 & 12 & -0.81 & 0.08 & -1.01 & 0.16 & 0.17 & 0.08 & 0.18 & 0.09 \\
PCII-14 &  64.8 &  6.6 & 187.0 & -29.5 & 2805 & 2826 &  65 & 353 & 1.99 & 12.3 & 1.7 &   44 & 31.3 & 17.6 & 14 & -0.82 & 0.07 & -1.15 & 0.13 & 0.13 & 0.05 & 0.32 & 0.08 \\
PCII-15 & 116.9 & 28.0 & 192.4 &  23.9 & 2055 & 2075 &  35 & 254 & 3.12 & 14.3 & 1.1 &   44 &  9.1 & 12.1 & 10 & -0.73 & 0.07 & -1.17 & 0.16 & 0.22 & 0.05 & 0.19 & 0.11 \\
PCII-16 & 139.4 & 30.4 & 195.6 &  43.5 & 2381 & 2401 & 114 & 527 & 3.50 &  7.6 & 2.0 & -103 & 13.0 & 10.1 & 11 & -1.36 & 0.09 & -1.71 & 0.27 & 0.27 & 0.09 & 0.21 & 0.09 \\
PCII-17 & 127.7 & 24.4 & 199.8 &  32.0 & 2315 & 2330 &  83 & 431 & 3.05 & 12.5 & 2.5 &  -40 & 18.1 & 17.7 & 10 & -1.46 & 0.15 & -1.34 & 0.25 & 0.15 & 0.11 & 0.21 & 0.11 \\
PCII-18 & 165.6 & 28.6 & 203.1 &  65.9 & 2855 & 2870 & 151 & 764 & 3.21 &  8.6 & 4.7 & -157 & 13.3 & 14.2 & 10 & -1.37 & 0.06 & -1.81 & 0.22 & 0.28 & 0.09 & 0.26 & 0.11 \\
PCII-19 & 139.9 & 22.2 & 206.6 &  41.9 & 2304 & 2319 & 102 & 609 & 2.72 & 14.5 & 3.4 &  -55 & 27.3 & 16.3 & 11 & -1.16 & 0.15 & -1.90 & 0.17 & 0.18 & 0.11 & 0.35 & 0.10 \\
PCII-20 & 132.6 &  6.1 & 221.5 &  29.2 & 2317 & 2332 &  69 & 470 & 3.04 & 10.9 & 1.6 &   71 & 13.2 & 11.4 & 15 & -0.96 & 0.09 & -1.63 & 0.22 & 0.19 & 0.06 & 0.18 & 0.08 \\
PCII-21\tablenotemark{a} & 134.0 &  3.2 & 225.2 &  29.0 & 2888 & 2913 &  74 & 514 & 3.50 & 10.3 & 1.4 &   85 & 14.9 &  4.0 & 17 & -1.02 & 0.05 & -1.44 & 0.12 & 0.12 & 0.05 & 0.26 & 0.08 \\
\enddata
\tablecomments{Kinematic and metallicity data for all class II peak
detection ECHOS from \citetalias{schl09}.  The columns are: right
ascension, declination, galactic longitude, galactic latitude, SEGUE
bright plate number, SEGUE faint plate number, number of SEGUE MPMSTO
spectra along the indicated line of sight, number of photometric MPMSTO
candidates along the indicated line of sight, volume scanned along the
indicated line of sight, median distance of the stars in the ECHOS, the
error in that estimate, the radial velocity of the ECHOS, the velocity
dispersion of the ECHOS, the median radial velocity error of the stars in
the ECHOS, the number of MPMSTO stars kinematically associated with the
ECHOS, the iron metallicity of the ECHOS, the error in that estimate,
the iron metallicity of the kinematically smooth component of the
halo along the indicated line of sight, the error in that estimate,
the $\alpha$-enhancement of the ECHOS, the error in that estimate, the
$\alpha$-enhancement of the kinematically smooth component of the halo
along the indicated line of sight, and the error in that estimate.}
\tablenotetext{a}{Monoceros Stream}
\tablenotetext{b}{\citet{gri06b} Stream}
\end{deluxetable}
\end{document}